\documentclass[aps,pre,reprint,superscriptaddress,showpacs,floatfix]{revtex4-1}
\usepackage{amsmath,amsfonts,amssymb,graphicx,natbib}
\usepackage{textcomp,mathcomp}
\usepackage{subfigure}
\usepackage{color,soul} 					  
\usepackage{comment}

\includecomment{comment}

\begin{document}


\title{Sharp interfaces in two dimensional free boundary problems: Efficient interface calculation via matched conformal maps}


\author{Stuart \surname{Kent}}
\affiliation{Program in Applied Mathematics, University of Arizona, Tucson, AZ 85721, USA}
\altaffiliation[Current address: ]{Detroit Labs, 1520 Woodward Ave., Suite 600, Detroit, MI 48226, USA}
\author{Shankar C.~\surname{Venkataramani}}
\affiliation{Program in Applied Mathematics, University of Arizona, Tucson, AZ 85721, USA}
\affiliation{Department of Mathematics, University of Arizona, Tucson, AZ 85721, USA}
\email[Author to whom correspondence should be addressed: ]{shankar@math.arizona.edu}

\date{\today}

\begin{abstract}
We use conformal maps to study a free boundary problem for a two-fluid  electromechanical system, where the interface between the fluids is determined by the combined effects of electrostatic forces, gravity and surface tension. The free boundary in our system develops sharp corners/singularities in certain parameter regimes, and this is an impediment to using existing ``single-scale" numerical conformal mapping methods.  The difficulty is due to the phenomenon of crowding, i.e.  the tendency of nodes in the preimage plane to concentrate near the sharp regions of the boundary, leaving the smooth regions of the boundary poorly resolved. A natural idea is to exploit  the scale separation between the sharp regions and smooth regions to solve for each region separately, and then stitch the solutions together. However, this is not straightforward as conformal maps are  rigid ``global" objects, and it is not obvious how one would patch two conformal maps together to obtain a new conformal map.  We develop a ``multi-scale" (i.e. adaptive) conformal mapping method that allows us to carry out this program of stitching conformal maps on different scales together. We successfully apply our method to the electromechanical model problem and discuss how it generalizes to other situations.
\end{abstract}

\maketitle


\section{Introduction}

Conformal mapping techniques are  powerful tools for analyzing two dimensional models, and have been applied to a variety of problems  \cite{schinzinger2012conformal} including aerodynamics \cite{milne1996}, fluid flows \cite{milne1996,Batchelor2000}, free boundary problems \cite{Jeong1992,Jeong2007,howison1992} potential theory/electrostatics \cite{tikhonov1990}, elasticity \cite{sokolnikoff1956} and even computer vision \cite{sharon20062d}. Abstractly, a problem of interest in the ``physical plane" is transformed into a simpler problem in the ``pre-image plane"  by an appropriately chosen complex analytic function, i.e. a conformal map \cite{nehari-conformal} relating the pre-image plane to the physical plane. In addition to generating analytic solutions, conformal mapping methods have also been applied in conjunction with numerical techniques to solve various problems of physical interest \cite{henrici1993}. 

An important issue in applying numerical (discretized) conformal maps to two dimensional problems where the domain has sharp regions/near singularities is the phenomenon of {\em crowding}. Crowding refers to the  tendency for image nodes to accumulate in the physical plane near regions of high boundary curvature and leave low curvature regions extremely coarsely represented. Crowding plagues all standard numerical conformal mapping methods when used to generate image domains with highly curved boundaries \cite{Porter2005,Wegmann2005}, 
and there are specialized algorithms to mitigate the effects of crowding for conformal mappings to {\em known} polygonal domains \cite{CRDT98}.
In this paper, we consider a model electrostatic free boundary problem where the interface can develop regions of extremely high curvature. We develop a matched asymptotics method for conformal maps,  that  completely avoids the computational difficulties associated with crowding for {\em a priori unknown domains}.

An interesting interplay between interface geometry and induced stresses exists in electromechanical systems. When a conducting 
interface is deflected by 
an external field, the induced charge distribution 
/the local electrostatic pressure is higher in the parts of the sheet with high curvature. This in turn causes even higher deflections and sharper curvatures. This  can produce extremely sharp and nearly singular shapes, as well as {\em the pull-in instability}, a runaway effect due to the positive feedback between curvature and electrostatic forces, whereby  no equilibrium solutions exist beyond a critical forcing strength. For these reasons we choose an electromechanical model as a test case for applying our techniques for multi-scale conformal maps. This model also has connections to recent studies that have been primarily driven by the development of microelectromechanical systems (MEMS), including micro-scale capacitors and actuators \cite{Bernstein2000,Chuang2010}.

The paper is organized as follows: In section~\ref{sec:model} we present the basic electromechanical model, a two fluid system where the interface is governed by a balance between electrostatic forces, gravity and surface tension. In section~\ref{sec:conformal} we present a  reformulation of our model problem in terms of conformal mappings, and discuss the advantages of such a reformulation. In section~\ref{sec:collocation} we present the basic numerical method for constructing numerical conformal maps, and apply it to situations where the free boundary is not ``sharp". The method breaks down as the interface develops sharp features, and in section~\ref{sec:multi_scale}  we present our matching technique for conformal maps that applies in a regime where the free boundary is sharp. We present a concluding discussion in section~\ref{sec:discussion}.

\subsection{Electromechanical model} \label{sec:model}

\begin{figure}[!ht]
\includegraphics[width=0.9 \linewidth]{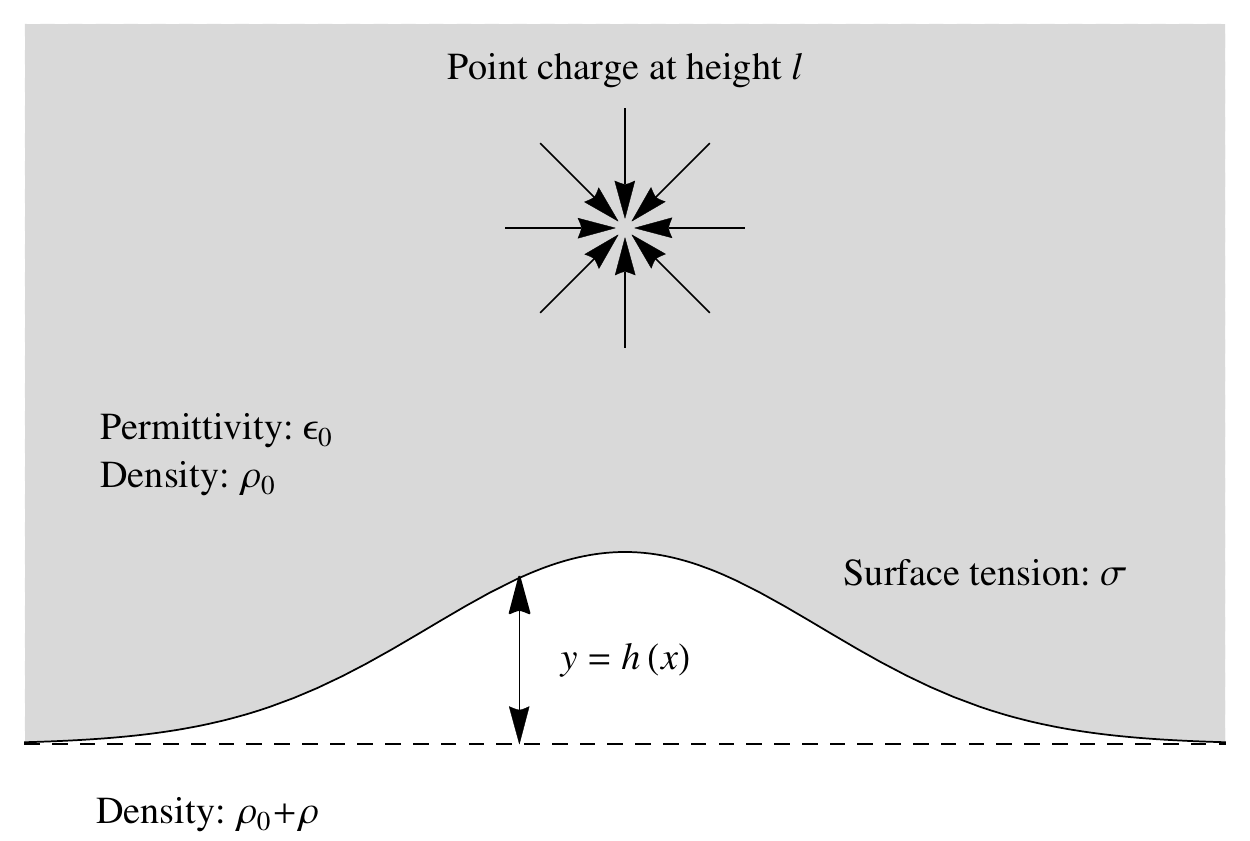}
\caption{\label{fig:generic_model}Generic model cross-section. The interface is between a dielectric and conducting fluid, and is gravitationally stable. $\mathbf{e}_x$ is in the horizontal direction, $\mathbf{e}_y$ is vertically upwards, and $\mathbf{e}_z$ is out of the page. The acceleration due to gravity is $-g \mathbf{e}_y$.}
\end{figure}

We consider a two fluid system -- an infinite conducting fluid initially in the half-space $y < 0$, and an infinite dielectric fluid in the half space $y > 0$. Gravity acts in the $-\mathbf{e}_y$ direction, and the density of the upper fluid $\rho_0$ is less than the density of the lower fluid $\rho_0 + \rho$, so the interface is gravitationally stable. The interface is deflected by a line charge with strength $q$ per unit length, along the $\mathbf{e}_z$ direction, placed in the upper fluid at a distance $l$ above the undeflected interface. Since all the induced charge on the fluid will reside on its surface, we can equally well consider a model with an infinite conducting sheet on an elastic foundation. The conducting sheet is initially at $y=0$ and is subsequently pulled up by a line charge placed at $y=l$. The restoring force from the foundation can be modelled as a linear function of the deflection from $y =0$, and this agrees with the expression for the hydrostatic pressure on the interface in the two-fluid model, which is also a linear function of the deflection. The setup is depicted in Fig.~\ref{fig:generic_model}, and by translation invariance in the $z$-direction, it suffices to consider a two dimensional problem in the $xy$ plane with a one dimensional interface $y = h(x)$.

The pressure balance along the deflected interface $y = h(x)$ is
\begin{equation*}
-\sigma \kappa(x) + \rho g h(x) = \frac{\epsilon_0}{2} \left(\frac{\partial \phi}{\partial n}\right)^2 \text{ for } x \in (-\infty,\infty).
\end{equation*}
where 
$\sigma$ is the surface tension of the interface, $\displaystyle{\kappa(x) = \frac{h_{xx}(x)}{(1+h_x^2(x))^{3/2}}}$ is the curvature of the interface at $(x,h(x))$, 
$\epsilon_0$ is the dielectric permittivity of the upper fluid, $\mathbf{n}$ is the unit normal to the interface, and $\displaystyle{\frac{\partial \phi}{\partial n} = \nabla \phi \cdot \mathbf{n} = |\nabla \phi|}$ is the magnitude of the electric field on the conducting interface. When forced by a line charge, the system is closed by the equations
\begin{align*}
\nabla^2 \phi(x,y) & = -\frac{q}{\epsilon_0}\delta(x,y-l) \text{ in } \lbrace h(x) < y : x \in (-\infty,\infty) \rbrace, \\ 
				 	  \phi(x,h(x)) & = 0 \text{ for } x \in (-\infty,\infty).
\end{align*}
The electric field $\nabla \phi$ vanishes as $x \to \pm \infty$, so we get $h(x) \to 0$ as $x \to \pm \infty$, {\em i.e} gravity stabilizes the interface for large $x$.
 
In order to simplify the equations, we pick units to make $\sigma = \rho g = \epsilon_0/2 = 1$. This is equivalent to the non-dimensionalization: $$
x' = \frac{x}{l_c}, y' = \frac{y}{l_c}, h' = \frac{h}{l_c}, l' = \frac{l}{l_c},
$$
for the lengths where $l_c = \sqrt{\sigma/\rho g}$ is the capillary length, and
$$
q' =  \sqrt{\frac{\epsilon^3_0}{2}}\left(\frac{\rho g}{\sigma^3}\right)^{1/4}q, \phi' = \sqrt{\frac{\epsilon_0}{2}}\left(\frac{\rho g}{\sigma^3}\right)^{1/4} \phi.
$$
We will henceforth work exclusively with the non-dimensionalized equations so there is no confusion in dropping the primes. This yields the nondimensional equations
\begin{subequations} \label{e:h_gov_pt}
\begin{align}
-\kappa(x) + h(x) & = |\nabla \phi(x,h)|^2 \text{ for } x \in (-\infty,\infty), \label{e:h_gov_pt_1} \\
\nabla^2 \phi(x,y) & = -q\delta(x,y-l) \text{ in } \lbrace y \geq h(x)   \rbrace , \label{e:h_gov_pt_2} \\ 
				 	  \phi(x,h(x)) & = 0 \text{ for } x \in (-\infty,\infty) \label{e:h_gov_pt_3} \\
					  h(x) \rightarrow 0 & \; \text{as} \; x \rightarrow \pm \infty. 
\end{align}
\end{subequations}
The system has a reflection symmetry $x \to -x$, and for simplicity we will restrict our attention to symmetric solutions satisfying $h(-x) = h(x), \phi(-x,y) = \phi(x,y)$.  The system has two dimensionless  parameters, a non-dimensional charge per unit length (i.e forcing) $q$ and a non-dimensional aspect ratio $l$, the ratio of the vertical length scale (the charge location) and the horizontal length scale, the capillary length. The local electrostatic pressure $\vert \nabla \phi \vert^2$ at each point on the sheet depends on the entire deflection profile $h(x)$ through the boundary conditions on $\phi$. This non-local coupling prevents the direct calculation of closed form solutions of \eqref{e:h_gov_pt}, and a numerical treatment is required instead.

\section{Conformal mapping reformulation} \label{sec:conformal}

We will now reformulate the system \eqref{e:h_gov_pt} in terms of conformal mappings, with the aim being to ``simplify" the non-local nature of the problem. $w = u+iv$ will denote the complex coordinate in the preimage plane and $z = x+iy$ the coordinate in the physical plane. 

\begin{figure*}[!ht]
\begin{center}
\subfigure[]
{
\includegraphics[width=0.45 \linewidth]{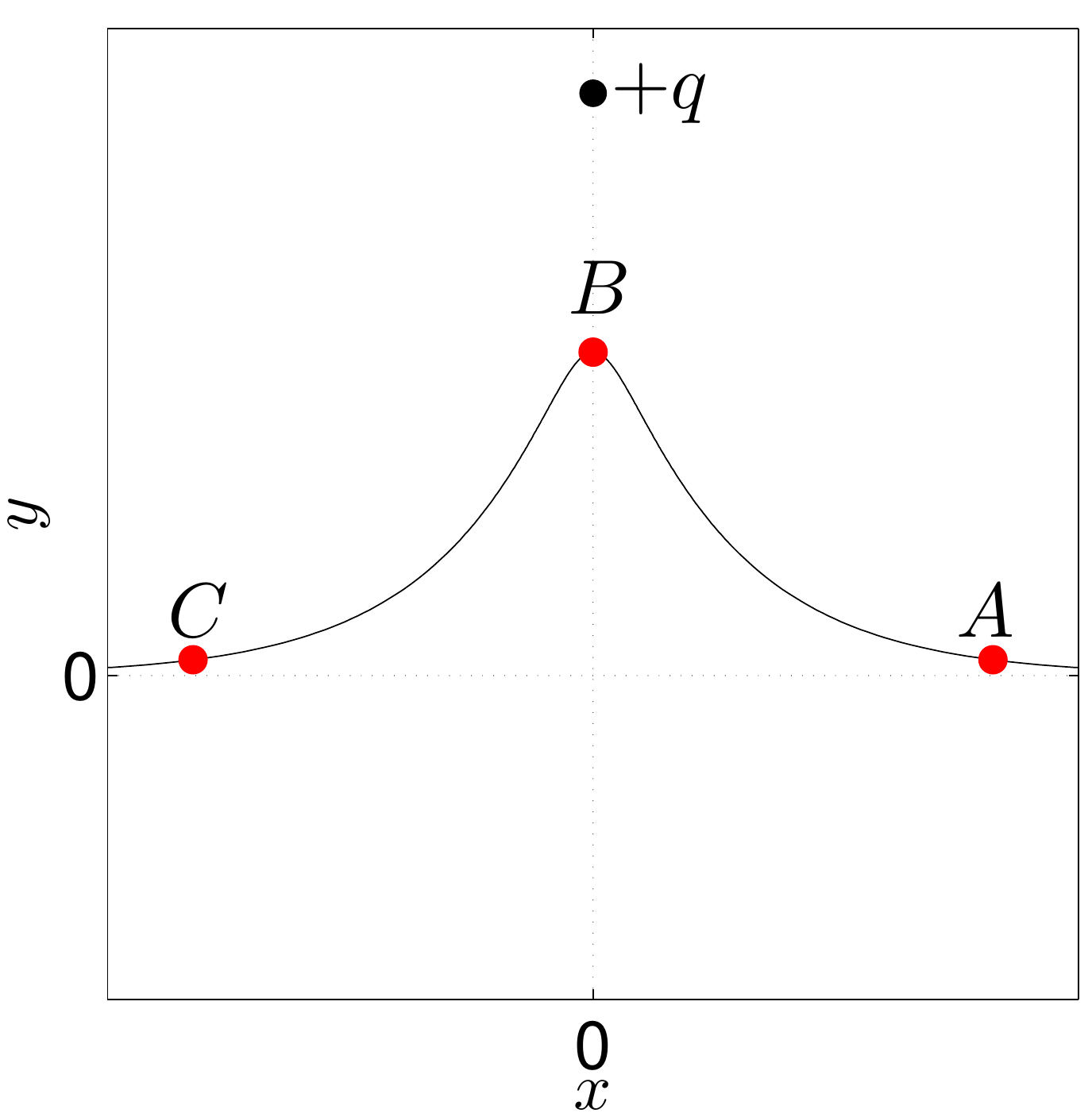}
\label{f:pt_full_symm_demo2}
}
\subfigure[]
{
\includegraphics[width=0.45 \linewidth]{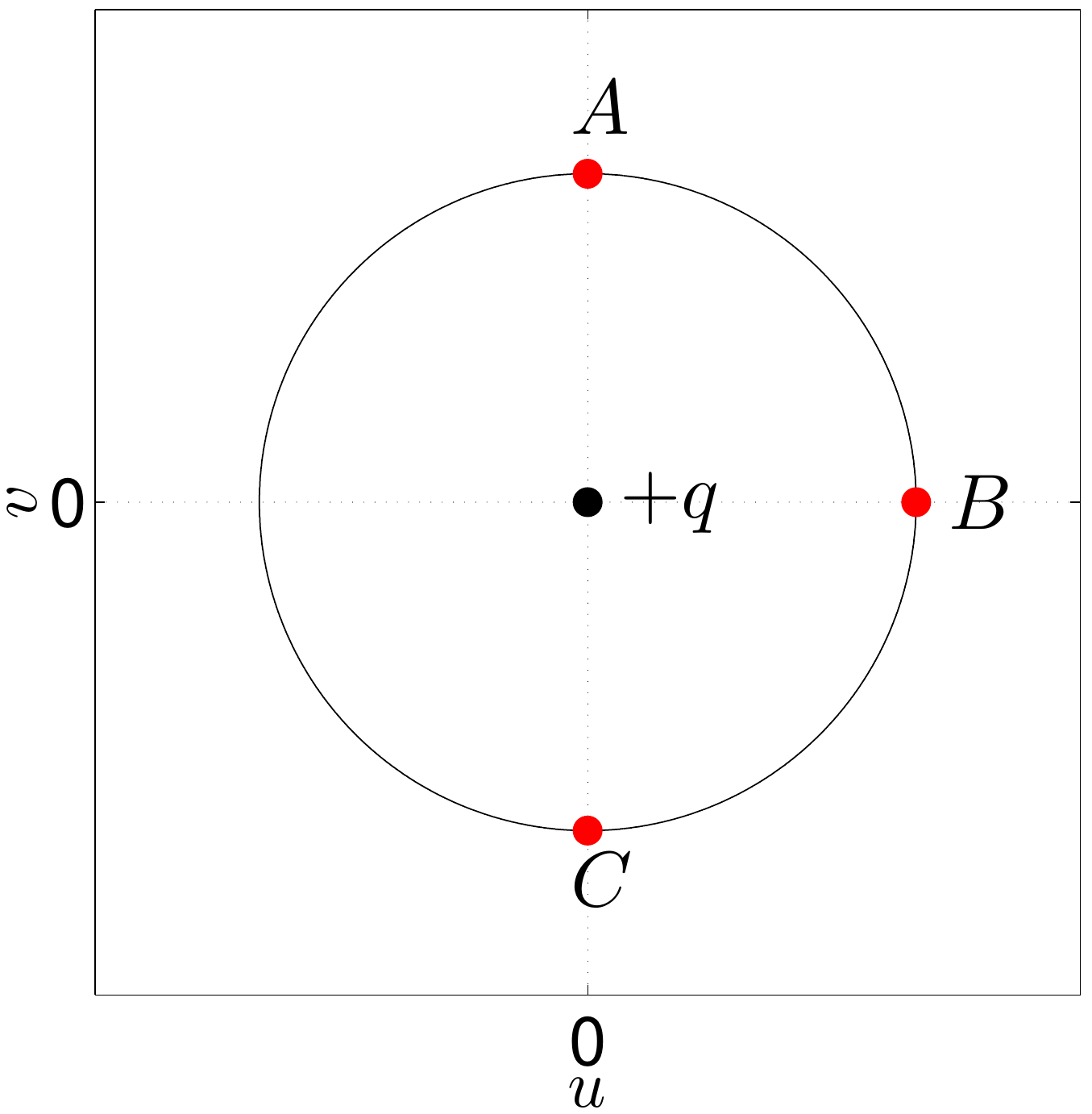}
\label{f:pt_full_symm_demo1}
}
\caption[Schematic diagrams showing candidate deflection profiles produced using conformal maps with appropriate symmetry conditions]{\label{fig:mapping} \subref{f:pt_full_symm_demo2} Example candidate deflection profile $z = x+iy = \lbrace if(\theta) : \theta \in (-\pi,\pi) \rbrace$ given by \eqref{eq:representation}. The deflection is symmetric about $x = 0$ and decays towards $y = 0$ as $\theta \rightarrow \pm \pi$. \subref{f:pt_full_symm_demo1} Unit circle in the preimage ($w = u+iv$) plane.  
}
\end{center}
\end{figure*}

Assume that we have a smooth interface $y = h(x)$ with $h(x) \to 0$ as $|x| \to \infty$. Then the upper fluid domain $\Omega_u = \{y > h(x)\}$ is a proper subset of $\mathbb{C}$. The Riemann Mapping theorem \cite{conway-book} thus asserts the existence of a bijective conformal map $z=F(w)$  such that $i F(D) = \Omega_u$, where $D$ is the (open) unit disk $|w| < 1$ in the preimage $w$-plane. This mapping is unique if we additionally impose the ``normalization" $i F(0) = i l$ and symmetry $F(\overline{w}) = \overline{F(w)}$ where the overbar denotes complex conjugation. The symmetry along with the unboundedness of $\Omega_u$ implies the  ``boundary condition"  $\lim_{w \to -1} |F(w)| = \infty$. The mapping $F$ is illustrated schematically in figure~\ref{fig:mapping}.

For the undeflected interface $h(x) = 0$, it is easy to check that the appropriate conformal mapping \cite{nehari-conformal} is 
$$
F_0(w) = l \frac{1-w}{1+w}
$$ 
For a deflected interface, assuming it is smooth and converges to the undeflected interface as $x \to \pm \infty$,  we show rigorously (mathematical details in a subsequent paper) that the conformal map $F$ is a ``bounded" modification of $F_0$ {\em i.e.}
\begin{equation}
F(w) = \alpha \frac{1-w}{1+w} + B(w)
\label{eq:representation}
\end{equation}
where $B$ is analytic on the {\em closed} unit disk $|w| \leq 1$, $\alpha$ is real, and we have the requirements
\begin{subequations}
\begin{align}
 B(\overline{w})  & = \overline{B(w)} & &\mbox{Symmetry}  \label{eq:symmetry}\\
 \alpha + B(0) & = l & & \mbox{Normalization} \label{eq:norm}\\
 B(-1) & = 0 && \mbox{Boundary condition} \label{eq:bc}
 \end{align}
 \end{subequations}
 The representation \eqref{eq:representation} shows that conformal map $F$ is continuous on $D \cup (\partial D \setminus\{-1\})$, i.e the only singularity of the map on the closed unit disk $|w| \leq 1$ is a pole at $w = -1$ with a {\em real} residue $2 \alpha$ corresponding to the boundary conditions $h(x) \to 0$ as $x \to \pm \infty$. The interface is then defined in terms of the conformal map $F$ parametrically by taking limits from inside the disk, i.e.:
$$
z = i f(\theta) \equiv \lim_{w_j \in D, w_j \to e^{i \theta}} i F(w) \mbox{ for }\theta \in (-\pi,\pi).
$$ 
This yields $z = i f(\theta) = \alpha \tan(\theta/2) + i B(e^{i \theta})$, a {\em conformal parameterization} of the the interface.

The invariance of Laplace's equation under conformal transformations now allows us to solve \eqref{e:h_gov_pt_2} using the electrostatic potential in the preimage plane to obtain
$$
\phi(z) = \frac{q}{2 \pi} \log|F^{-1}(z)|.
$$ 
A direct calculation using the parameterization $z = if(\theta) = x(\theta) + i h(x(\theta))$ so that $f(\theta) = h(x(\theta))-ix(\theta)$ shows that 
\begin{align*}
h(x(\theta)) & = \text{Re}[f(\theta)] \\
\kappa(x(\theta)) & = \frac{x'(\theta) h''(\theta) - h'(\theta) x''(\theta)}{(x'(\theta)^2 + h'(\theta)^2)^{3/2}} = \frac{\text{Im}[f_{\theta\theta}(\theta)\overline{f_{\theta}}(\theta)]}{\vert f_{\theta}(\theta) \vert^3}
\end{align*}
The force balance \eqref{e:h_gov_pt_1} thus reduces to a single equation for $f$:
\begin{equation}
\frac{q^2}{4 \pi^2 \vert f_{\theta}(\theta) \vert^2} - \text{Re}[f(\theta)] + \frac{\text{Im}[f_{\theta\theta}(\theta)\overline{f_{\theta}}(\theta)]}{\vert f_{\theta}(\theta) \vert^3} = 0. \label{eq:new_ode}
\end{equation}
Observe that this equation is {\em local} in $\theta$ unlike the original formulation \eqref{e:h_gov_pt} where the force balance equation is {\em nonlocal}. Also observe that we have one ``real" equation for a ``complex" function $f$, so we should expect additional condition(s) that determine $f$.

For any function $\Upsilon$ that is analytic on a domain containing the closed unit disk, the real and imaginary parts of $\upsilon(\theta) = \Upsilon(e^{i \theta})$ are not independent. Rather, they are related by the periodic Hilbert transform \cite{garling-inequalities} 
\begin{align*}
\text{Re}[\upsilon] & = -\mathcal{H}[\text{Im}[\upsilon]] + \text{Re}[\Upsilon[0]] \\
\text{Im}[\upsilon] & = \mathcal{H}[\text{Re}[\upsilon]] + \text{Im}[\Upsilon[0]]
\end{align*}
because they are boundary values of conjugate Harmonic functions on the unit disk \cite{conway-book}. The periodic Hilbert transform $\mathcal{H}$ of a periodic function $\xi$ is defined \cite[\S 11.5]{garling-inequalities} by
$$
\mathcal{H}[\xi](\phi) = \lim_{\epsilon \to 0} \frac{1}{2\pi}\int_{\epsilon}^\pi \cot\left(\frac{\theta}{2}\right) \left[\xi(\phi-\theta) -\xi(\phi+\theta)\right] d\theta 
$$
We will rewrite this condition in a slightly different form that exploits the  natural mapping between analytic functions on the closed unit disk and Fourier series given by 
$$
\Upsilon(w) = \sum_n a_n w^n  \leftrightarrow
 \sum_n a_n e^{i n \theta} = \upsilon(\theta)
 $$
where the sums do indeed converge because the coefficients $a_n$ decay exponentially, and $\upsilon(\theta) = \Upsilon(e^{i \theta})$. Since $\Upsilon$ is analytic at 0, it follows that 
\begin{equation}
a_{-n} = \frac{1}{2 \pi} \int_0^{2 \pi} \upsilon(\theta) e^{i n \theta} d \theta = 0 \mbox{ for } n = 1,2,\ldots
\label{eq:constraint}
\end{equation}
These equations are {\em local} in the transform domain (in terms of $n$), but are {\em nonlocal} integral equations in terms of $\theta$.  From \eqref{eq:representation}, it follows that $\Upsilon(w) = (1+w) F(w)$ is analytic on the closed unit disk, so choosing
$$
\upsilon(\theta) = \Upsilon(e^{i \theta}) = (1+e^{i \theta}) f(\theta),
$$ 
in \eqref{eq:constraint} yields
\begin{equation}
\int_0^{2 \pi} (1+e^{i \theta}) f(\theta) e^{i n \theta} d \theta = 0 \mbox{ for } n = 1,2,\ldots
\label{eq:hilbert}
\end{equation}
The symmetry condition \eqref{eq:symmetry} implies 
\begin{equation}
f(-\theta) = \overline{f(\theta)} \mbox{ for } \theta \in (-\pi,\pi) 
\label{f_symmetry}
\end{equation}
Finally, we remark that the normalization \eqref{eq:norm} is {\em nonlocal} in terms of $f$ since 
\begin{equation}
l = F(0) = \Upsilon(0) = \frac{1}{2 \pi} \int_0^{2 \pi} (1+e^{i \theta})f(\theta) d \theta, 
\label{f_norm}
\end{equation}
and the boundary condition \eqref{eq:bc} can be recast as 
\begin{equation}
\lim_{\theta \to \pm \pi} f(\theta) = 0 \mp i \infty \label{f_bc}
\end{equation}
Equations \eqref{eq:new_ode},\eqref{eq:hilbert}--\eqref{f_bc} together give a reformulation of \eqref{e:h_gov_pt} in terms of a (normalized) conformal parameterization of the interface. The key advantage of this reformulation is that, every equation is local either on the unit circle in the preimage plane (i.e. in terms of $\theta$) or in the Fourier transform domain (i.e. in terms of $n$). Since we can efficiently switch between functions on the unit circle and their Fourier transforms using FFTs \cite{num-recipies}, this reformulation, if discretized appropriately, will give an efficient numerical algorithm for solving \eqref{e:h_gov_pt}.

\section{Collocation method}
\label{sec:collocation}

To solve for the interface numerically, we will discretize the representation for $F$ in \eqref{eq:representation}. For clarity we use a tilde to denote discretized numerical approximations, {\em e.g.} $\widetilde{F}$ is a numerical approximation to $F$. Since every analytic function on the unit disk has a convergent Taylor series \cite{conway-book}, we can define families of conformal maps with finitely many degrees of freedom by truncating the power series for $B$ centered at 0:
\begin{equation}
\widetilde{F}(w) = \alpha \left(\frac{1-w}{1+w}\right) + \sum_{j=0}^{M} \beta_j w^j.
\label{eq:discretize}
\end{equation}
In our numerics we always take $M = 2^K$, a power of 2. Then $\widetilde{f}(\theta)$ given by restricting $\widetilde{F}$ to the unit circle automatically satisfies \eqref{eq:hilbert}. Explicitly
\begin{equation}
\widetilde{f}(\theta) = -i \alpha \tan \left(\frac{\theta}{2}\right) + \sum_{j=0}^{M} \beta_j e^{i j \theta} \mbox{ for } \theta \in (-\pi,\pi). \label{eq:full_maps}
\end{equation}
The symmetry requirement \eqref{f_symmetry} implies that $\alpha, \beta_0,\beta_1,\ldots,\beta_{M}$ are all real, so the family of conformal maps in \eqref{eq:full_maps} satisfying \eqref{eq:hilbert} and \eqref{f_symmetry} is parameterized by $M+2$ real quantities. The normalization \eqref{eq:norm} yields
\begin{equation}
l = \alpha + \beta_0 \label{at0}
\end{equation}
and the boundary condition \eqref{eq:bc} implies that
\begin{equation}
\sum_{j=0}^M (-1)^j \beta_j = 0 \label{at-1}
\end{equation}
We now turn to discretizing \eqref{eq:new_ode}. Let us first consider the problem of determining $\widetilde{f}$ given $q$ and $l$. In general, we cannot expect the (numerical) residual 
\begin{equation}
\displaystyle{R[\widetilde{f}] \equiv \frac{q^2}{4 \pi^2 \vert \widetilde{f}_{\theta}(\theta) \vert^2} - \text{Re}[\widetilde{f}(\theta)] + \frac{\text{Im}[\widetilde{f}_{\theta\theta}(\theta)\overline{\widetilde{f}_{\theta}}(\theta)]}{\vert \widetilde{f}_{\theta}(\theta) \vert^3}}
\label{eq:residual}
\end{equation} to vanish for all $\theta \in (-\pi,\pi)$ for our numerical maps with finitely many degrees of freedom. 

The residuals at points $\theta$ and $-\theta$ are {\em not independent} since $\widetilde{f}(-\theta) = \overline{\widetilde{f}(\theta)}$ so that $R[\widetilde{f}](\theta) = R[\widetilde{f}](-\theta)$. Consequently, we get a possible discretization of \eqref{eq:new_ode} for  every choice of $M$ distinct points $\lbrace\theta_m:m=0,1,\ldots,M-1\rbrace $ such that $\theta_m \neq -\theta_n$ for any pair $m \neq n$, Without loss of generality, we can take these $M$ points in $[0,\pi)$, i.e with non-negative angles.  As we discussed in Sec.~\ref{sec:conformal}, we will get efficient numerical algorithms if we can easily switch between values of the function $f(\theta_m)$ and its Fourier coefficients $a_n$. This can be achieved using FFTs if we pick the points $\theta_m$ uniformly spaced on the unit circle and $M = 2^K$. In particular, for the discretization in \eqref{eq:full_maps}, we can apply the FFT to compute the function values and derivatives at evenly spaced nodes $\theta_m = m\pi/M$, $m = 0,\ldots,M-1$.

The quantities $\alpha$ and $\beta_j, j = 0,1,2,\ldots,M$ are now determined from \eqref{at0},~\eqref{at-1}~and the conditions~$R[\tilde{f}](\theta_m) = 0$ for $m=0,1,2,\ldots,M-1$.  In implementing this method, we find that the numerical algorithm is better conditioned if we instead parameterize the maps by the $M+2$ quantities $\alpha$ and $h_m \equiv h(\theta_m) = \text{Re}[f(\theta_m)]$.In this parameterization the boundary condition \eqref{eq:bc} implies that $h_M = 0$, so we have $M+1$ independent real parameters. We also have $M+1$ real conditions from the vanishing of the residual at $\theta_m, m=0,1,2,\ldots,M-1$ and the normalization \eqref{at0}. The coefficients $\beta_j$ and the ``height" parameters $h_0,h_1,\ldots,h_{M-1}$ are of course related by a linear transformation which can be expressed in terms of the Discrete Fourier transform.

The final point to consider is that we {\em do not expect} the system \eqref{e:h_gov_pt} will have a unique global solution for all given $q$ and $l$, since we expect the system to display multiple equilibrium solutions for some choices of $q$ and $l$ based on the analogy with the MEMS model \cite{Pelesko2002}. As in the analysis of the MEMS model, we use a continuation method \cite{Pelesko2002} with the tip  height of the interface, $h_0$ and the charge location $l$ as the specified parameters. $l$ and $h_0$ are thus the independent quantities, and the equilibrium equations determine  the charge $q$ and the other heights $h_1,h_2,\ldots,h_{M-1}$ as functions of $h_0$ and $l$. Given $l$ and $h_0$, we compute the residual $R[\widetilde{f}](\theta_m)$ as a function of $(q,h_1,h_2,\ldots,h_{M-1})$ through the following steps:
\begin{enumerate}
\item Compute the Fourier coefficients $\beta_j$ corresponding to the height parameters $h_m$ where $h_M = 0$.
\item Compute $\widetilde{f}(\theta_m),\widetilde{f}_\theta(\theta_m)$ and $\widetilde{f}_{\theta\theta}(\theta_m)$ using the Inverse FFT and the representations $\widetilde{f}(\theta) = \sum  \beta_j e^{i j \theta}, \widetilde{f}_\theta(\theta) = \sum  i j \beta_j e^{i j \theta}$ and $\widetilde{f}_{\theta \theta}(\theta) = \sum  - j^2 \beta_j e^{i j \theta}$.
\item Compute the residual $R[\widetilde{f}]$ at the points $\theta_m, m = 0,1,2,\ldots,M-1$ using \eqref{eq:residual}.
\end{enumerate}
We then compute a numerical approximation of the true deflection profile and corresponding charge strength $q$ by minimizing $\sum_m (R[\widetilde{f}](\theta_m))^2$ using a nonlinear least squares method implemented in the MATLAB function {\tt lsqnonlin} \cite{matlab}. Note that every (local) minimum for the sum of the squares of the residual is {\em not necessarily an approximate solution} to \eqref{e:h_gov_pt}. We in fact need that the {\em residual vector be zero} to within a prescribed tolerance.

Figure \ref{fig:orig_bifn_diags} shows examples of bifurcation diagrams computed using this method for $l \in \lbrace 0.25,0.5,0.75,1 \rbrace$. For all  these choices of $l$, the $h(0)$ vs. $q$  curves exhibit saddle-node bifurcations  i.e. they ``turn around" at a critical value of $q^*(l)$. For values of $q > q^*(l)$, the system \eqref{e:h_gov_pt} has {\em no solution}. Physically, if we perform the experiment of increasing $q$ quasi-statically from 0 past the critical value $q^*$, the tip height $h_0(q)$ will track the lower branch of the appropriate curve in  Figure~\ref{fig:orig_bifn_diags}, until $q = q^*$ at which point the conducting fluid will be drawn to the point charge ``grounding" it.  This is the the analog in the two-fluid system of the  pull-in instability at finite forcing strength $q$ as reported for small aspect ratio MEMS \cite{Bernstein2000}. 

Deflections below and above the saddle-node bifurcation are stable and unstable respectively depending on the sign of $\displaystyle{\frac{d h_0}{d q}}$. Indeed, along the upper branch, a small perturbation of $h_0$ away from the equilibrium value $h_0(q)$ will result in an interface that is subject to a higher forcing than necessary to support it (if the deflection is toward the charge) or a lower forcing than necessary (if the deflection is away from the charge). In either case, the interface will evolve dynamically and either be drawn to the point charge or relax to the (stable) equilibrium on the lower branch, so the upper branch of equilibria is unstable to small perturbations. 

The solid dots on the figure represent the largest $h_0$ (for each $l$) for which the method converges with $M = 256$. While this limiting $h_0$ increases with $M$, it does so very slowly and we cannot get much closer to the vertical axis for any value of $M \lesssim 4096$ (the bound on $M$ reflects our limitations on memory size and processor speed). Nonetheless, the figure is very suggestive that, for all $l$, the bifurcation curve approaches the point $(q=0,h_0=l)$ (the solid dots on the vertical axis).

\section{Multiple scale analysis} \label{sec:multi_scale}

\begin{figure*}[!ht]
\begin{center}
\subfigure
{
\includegraphics[width=0.45 \linewidth]{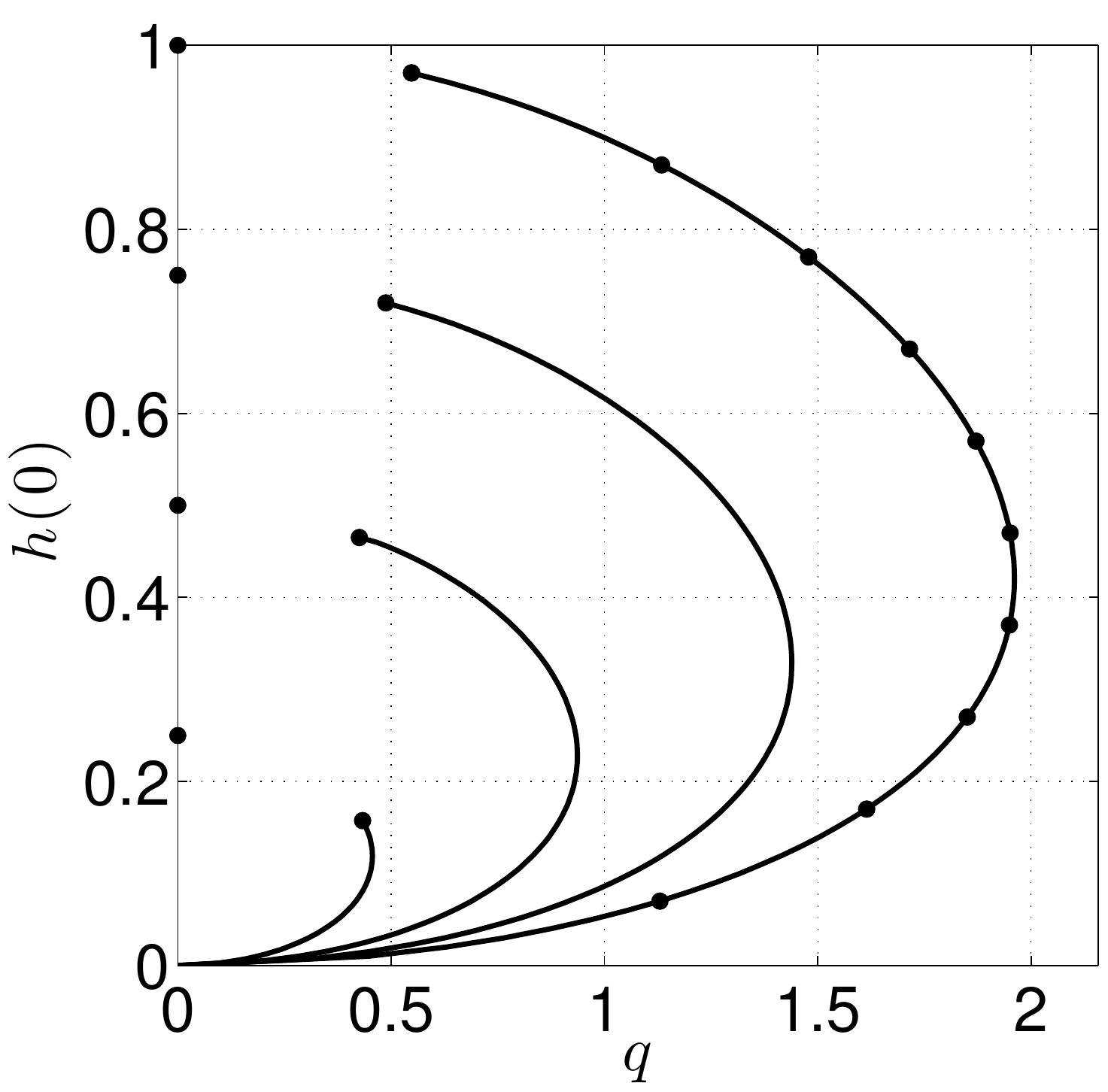}
\label{fig:orig_bifn_diags}
}
\subfigure
{
\includegraphics[width=0.45 \linewidth]{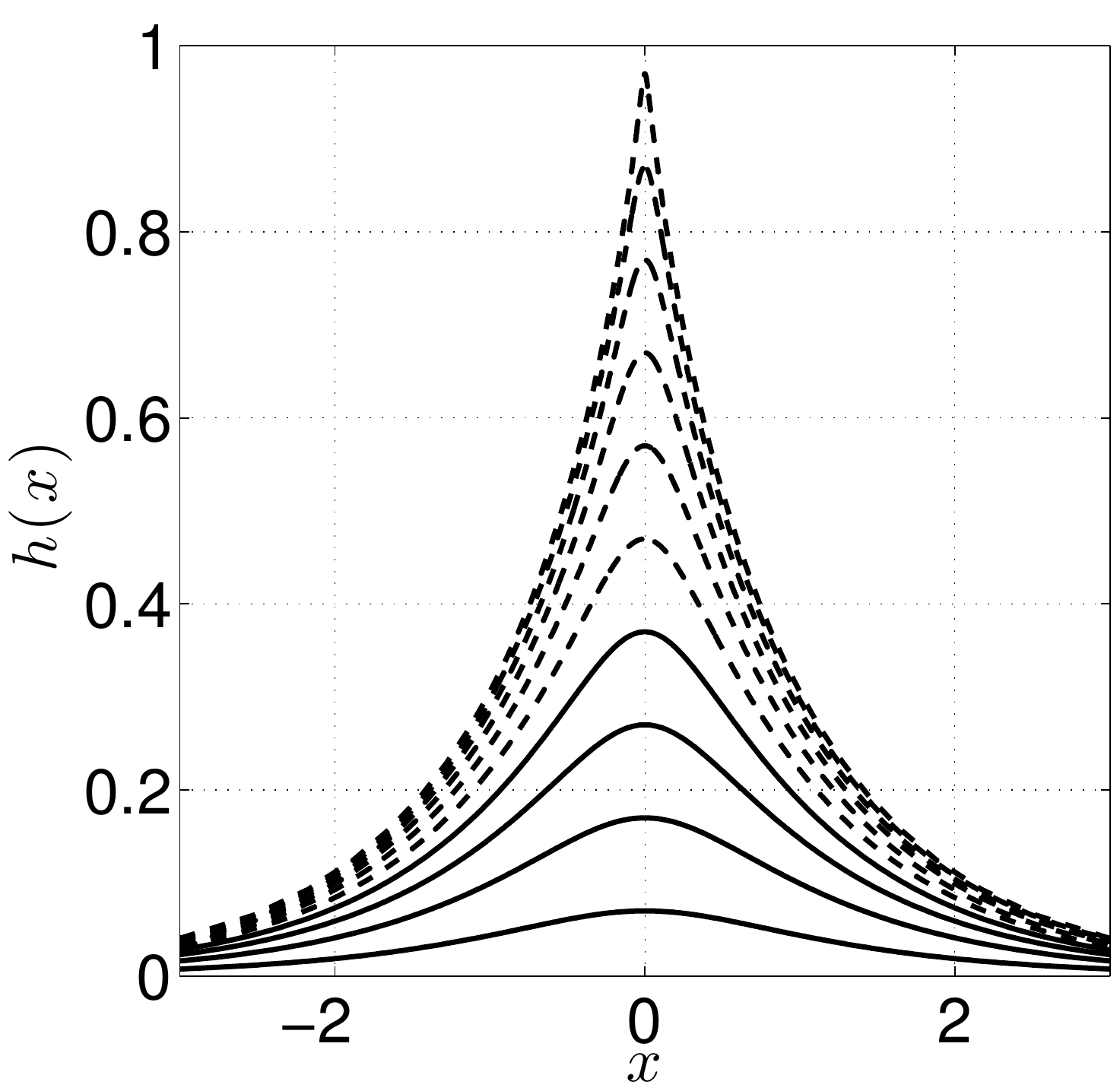}
\label{fig:orig_solutions}
}
\caption{\subref{fig:orig_bifn_diags} Computed bifurcation diagrams (sheet tip height $h(0)$ plotted against charge strength $q$) for charge heights $l$ = 0.25, 0.5, 0.75, 1 (left to right). All the curves have a maximum value of $q = q^*(l)$ beyond which there are no solutions. This indicates the existence of a maximally-deflected stable solution at a critical forcing strength, numerically demonstrating a {\em saddle-node} bifurcation for this system. The upper branch of the bifurcation curves correspond to (dynamically) unstable profiles because $h_0'(q) < 0$. There is also a (independent) critical deflection magnitude, the solid dots at the end of each curve, beyond which the numerical scheme used breaks down due to crowding. $M = 256$ for all the curves. The solid dots on the vertical axis represent the charge locations, and it is suggestive that, if continued, the bifurcation curves will converge to these points. \subref{fig:orig_solutions}
Numerically computed solutions $h(x)$ with the charge location $l=1$ and $h_0$ varied. Each profile corresponds to a single solid dot on the $l=1$ (rightmost) curve in \subref{fig:orig_bifn_diags}. The solutions were generated using the collocation method with $M=256$. The dashed curves and are unstable profiles and the solid curves below  are stable. For $l=1$, the maximally deflected profile that we can compute with the collocation method corresponds to $h_0 = 0.96$. }
\end{center}
\end{figure*}

Figure~\ref{fig:orig_solutions} shows the numerically computed solutions for different tip heights $h_0$ with $l$ fixed at 1. The figure shows both stable ($h_0 \lesssim 0.45$) and unstable   profiles. The highest profile that we can compute before the collocation method breaks down with $M = 256$ is $h_0 \approx 0.96$.
It is evident from the figure that the tips (region near $x = 0$) are getting very sharp on the scale of the capillary length (nondimensionalized to 1) as $h_0$ approaches $l$. This reflects {\em non-uniformity/multiple scale behavior} of the solutions of \eqref{e:h_gov_pt}. In particular, the small scale structures are governed by a different balance (surface tension vs electrostatic pressure) from the large scale structure (surface tension vs gravity). A natural approach that will accounts for this non-uniformity it to discretize the problem {\em adaptively}, i.e. in a manner which reflects the local length scale in the solution. One would like to  have roughly equal number of points to resolve each of the regions with a different dominant balance.

\begin{figure}[!ht]
\includegraphics[width=0.9 \linewidth]{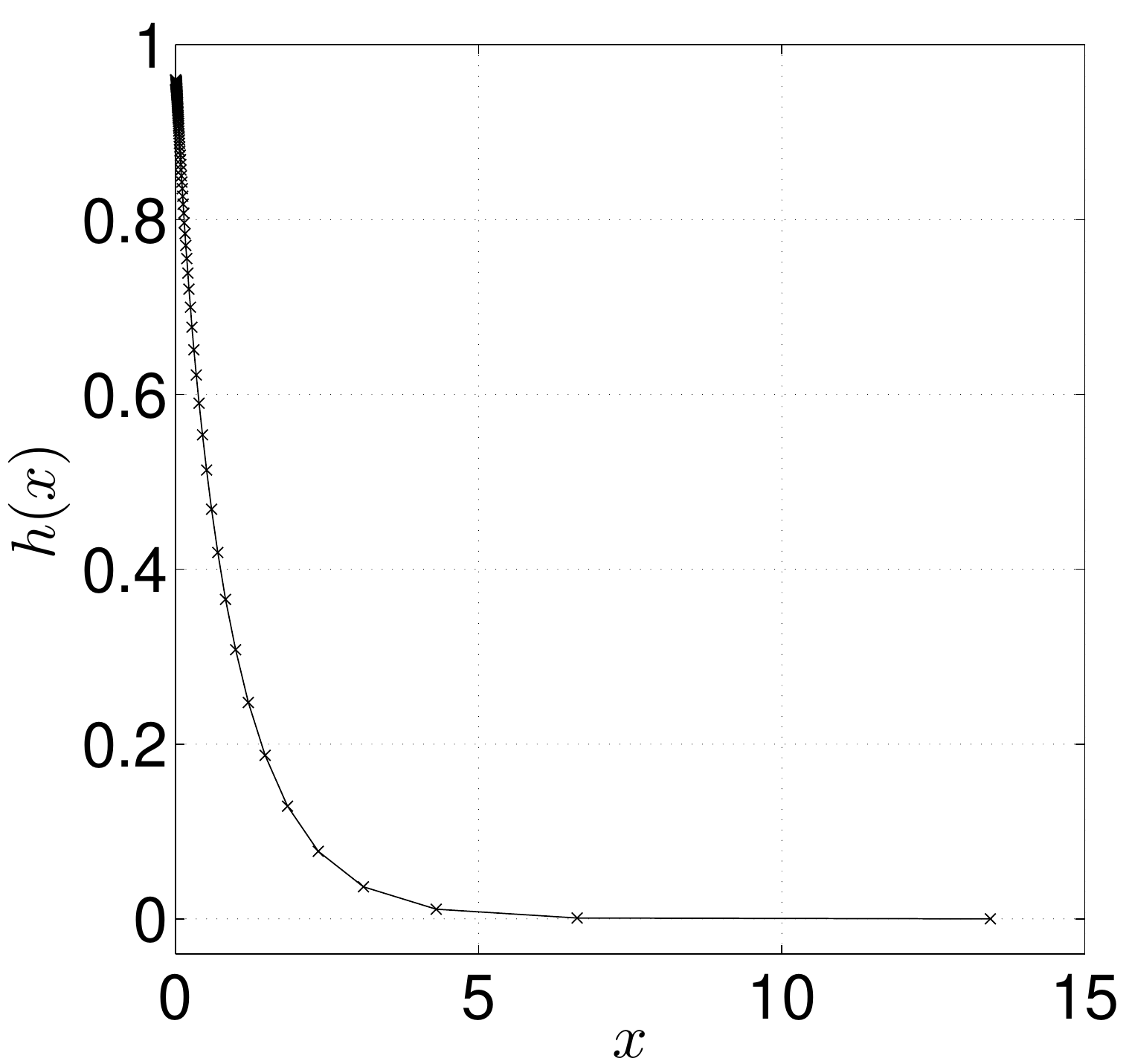}
\caption{\label{f:nodes} Deflection profile computed with $l = 1$, $h(0) = 0.96$, $M = 256$. The density of the image nodes $\lbrace i f(i j \pi/2M) : j = 0,\ldots,M-1 \rbrace$ decreases rapidly away from the interface tip. This leaves the decaying portion of the interface poorly resolved. The bulk of the nodes are in the near-tip region which is extremely well-resolved.}
\end{figure}

A weakness of the conformal mapping method, as described in Section~\ref{sec:collocation} is that it is intrinsically {\em non adaptive}. We do not get to pick the points $\theta_m$ which are the ``nodes" in our discretization of the continuous system. Rather, they are required to be uniformly spaced on the preimage plane in order to use FFT techniques. Figure~\ref{f:nodes} 
shows the locations in the physical plane (i.e. on the interface) of uniformly space nodes on the unit circle for $l = 1, h_0 = 0.96, M = 256$. Contrary to our initial expectation, the breakdown of the collocation method is not due to poor resolution of the tip; rather it is due to poor resolution of the decaying region of the interface, where the dominant ``large-scale" balance is between gravity and surface tension, and the relevant length scale for this region is the capillary length. 

This effect is easily understood by recognizing that the electric field on the interface is related to the inter-node separation in the physical plane since $|\nabla \phi| \sim 1/|f_\theta| \sim (\text{node separation})^{-1}$. Since the electric field (and the induced charge) concentrates near sharp tips, the nodes in the physical plane will concentrate near sharp tips. The tendency for image nodes to accumulate near regions of relatively high interface curvature, known as \emph{crowding}, is ubiquitous when using conformal maps defined on the unit disk $D$ to represent interfaces that feature very disparate length scales.  It has been recognized as a substantial impediment to using numerical methods to find conformal maps for regions that have multiple length-scales/singularities in their boundary \cite{Wegmann2005,Porter2005}. 

This is not a peculiarity of the conformal mapping based method: alternative methods that we could have applied, including boundary element methods based on the boundary integral formulations, or direct minimization of the total system energy, must all be adapted to deal with the disparate length scales that characterize stiff problems. Another approach to avoid crowding is to pick the nodes adaptively (and possibly non-uniformly) in the preimage plane. In this case, FFT methods do not apply. The Hilbert transform constraint relating the real and the imaginary parts of the interface parameterization $f$ yields a linear system with a matrix that has a poor condition number \cite{Wilkening2011}. This constraint has to be imposed along with the minimization of the residual \eqref{eq:residual}, and Wilkening \cite{Wilkening2011} has applied this idea with careful and sophisticated numerical methods to analyze the crests of large amplitude standing water waves, which were conjectured to have a corner singularity in the free interface \cite{PenneyPrice1952}.  In section~\ref{sec:matching} below, we detail a different approach to this problem, one that borrows from the philosophy of matched asymptotics in multiple scale analysis \cite{Hinch1991}. 

\subsection{Concentrated charge approximation} \label{sec:concentrated}

The collocation method breaks down when the tip of the interface $h_0$ approaches the charge location $l$. This motivates the definition $\epsilon = l-h_0$, as the relevant small parameter that defines this asymptotic regime. When $\epsilon$ is small, we expect that the tip is sharp and the total charge $-q$ induced on the sheet accumulates (and hence, image nodes crowd) near the high-curvature sheet tip. 

As $\epsilon \to 0$, all the charge concentrates on the tip, so to leading order in $\epsilon$, we can model the charge on the interface as a concentrated charge $-q$ at the tip $(0,h_0)$. 
The interface is then effectively supported by an upward electrostatic point force of strength 
\begin{equation}
\Lambda \approx q^2/2\pi(l-h_0) \label{force}
\end{equation} 
applied to the tip, and relaxes under the influences of gravity and elasticity everywhere else (where the electric field/induced charge is zero in this leading order approximation). The resulting leading order deflection profiles are governed by a collapsed form of the  equation \eqref{e:h_gov_pt_1},
\begin{equation}
\Lambda \delta(x) - h_{out}(x) + \frac{h''_{out}(x)}{(1+(h'_{out}(x))^2)^{3/2}} = 0, \label{elastic}
\end{equation}
which can be integrated exactly. 
Multiplying \eqref{elastic} by $h_{out}'$ and integrating in from infinity yields
\begin{equation}
\frac{1}{\sqrt{1 + (h'_{out}(x))^2}} + \frac{(h_{out}(x))^2}{2} = 1, \label{e:pt_out_ode_1st_int}
\end{equation}
where the constant term on the right hand side has been determined by applying the asymptotic boundary condition. We can solve this equation, provided that $0 \leq h_{out}(x) < \sqrt{2}$,  to get the implicit representation
\begin{widetext}
\begin{equation}
\log{\left(\frac{2+\sqrt{4-(h_{out}(x))^2}}{2+\sqrt{4-(h_{out}(0))^2}}\right)} - \log{\left(\frac{h_{out}(x)}{h_{out}(0)}\right)} + \sqrt{4-(h_{out}(0))^2} - \sqrt{4-(h_{out}(x))^2} = |x|
\label{eq:outer-profile}
\end{equation}
\end{widetext}
For $0 \leq h_{out}(x) \leq h_{out}(0) < \sqrt{2}$, we see that the first, third and fourth terms in the above expression are bounded and $O(1)$. Consequently, for large $|x|$, we have 
\begin{equation}
h_{out}(x) \simeq h_{out}(0)e^{-|x|}
\label{eq:outer-outer-profile}
\end{equation}
Integrating \eqref{elastic} across $x = 0$, we see that  $h_{out}$ is continuous but has a corner (discontinuous slope) at $x=0$:
\begin{equation}
\frac{h'_{out}(0^+)}{\sqrt{1+(h'_{out}(0^+))^2}}-\frac{h'_{out}(0^-)}{\sqrt{1+(h'_{out}(0^-))^2}} = - \Lambda,
\label{eq:jump}
\end{equation}
where $\Lambda$ is the total force on the interface from the charge (or equivalently, by Newton's third law, the force on the charge pulling it towards the interface). This equation is a statement of the balance between the surface tension forces on a corner and the total electrostatic force on the interface.

\begin{figure}[!ht]
\begin{center}
\includegraphics[width=\linewidth]{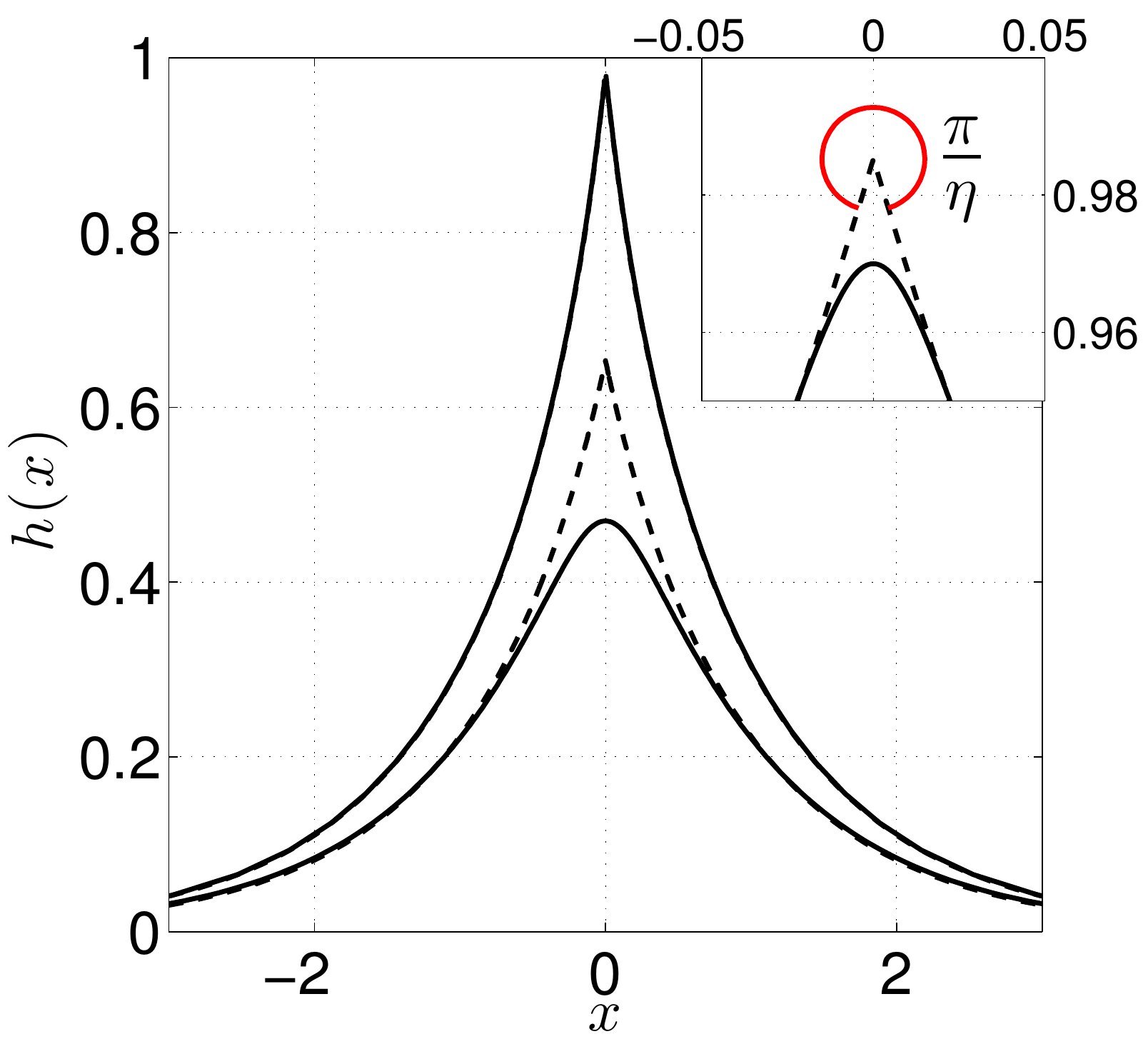}
\end{center}
\caption[Leading order large deflection profiles plotted against full deflection profiles] {\label{fig:schematic_solns}Leading order large deflection profiles ($h_{out}(0) = 0.985$, $h_{out}(0) = 0.65$, dashed curves) corresponding to computed interfaces from Fig.~\ref{fig:orig_solutions} ($h_0 = 0.97$. $h_0 = 0.47$, solid curves). The inset shows a zoomed view near the point $(0,1)$. The width of the region in which the leading order large deflection approximation is inaccurate approaches zero as $\epsilon \rightarrow 0$ and the outer tip angle is $\displaystyle{\frac{\pi}{\eta}}$.}
\end{figure}

We will call the solutions in \eqref{eq:outer-profile} the leading order, large deflection solutions, because they are the appropriate limit solutions as $\epsilon \to 0$ (tip curvature $\to \infty$) and they cannot be obtained as a perturbation expansion in powers of $q$ about the $q=0, h = 0$ undeformed interface. In particular, they are not smooth and have a corner at $x=0$, unlike the perturbation series solutions in powers of $q$ which are necessarily smooth.  We define a parameter $\eta=\eta(h_{out}(0))$ such that $\pi/\eta(h_{out}(0))$ denotes the outside angle of the leading order interface tip corner, so that $1/2 \leq \eta \leq 1$ (see Fig.~\ref{fig:schematic_solns}). Eq.~\eqref{e:pt_out_ode_1st_int} 
now yields
\begin{equation}
\eta = \left(2 - \frac{2}{\pi}\arctan{\left( \frac{2-(h_{out}(0))^2}{h_{out}(0)\sqrt{4-(h_{out}(0))^2}} \right)}\right)^{-1},
\label{eq:gamma}
\end{equation}
where $\arctan$ takes values in $[0,\pi/2]$.

Figure~\ref{fig:schematic_solns} shows the comparison between the solutions in \eqref{eq:outer-profile} and the numerically computed profile $h(x)$ from the collocation method with $l=1$. Note that the solutions agree very well outside the tip region, but in the tip region, there are consistent deviations between the leading order, large deflection solution and the numerical solution of the interface. Electrostatic forces are important near the tip, and our crude approximation of the electrostatic force is insufficient to resolve the tip region. 

An important point to note is that $h_{out}(0) \neq h(0) = h_0$. This brings up the question of how to determine the appropriate value of $h_{out}(0)$ that corresponds to a given $h_0$ with the charge location fixed at $l$. We address this question in more detail below, but the basic idea is that the ``right" parameters to describe the sharp-tip asymptotic regime are $h_{out}(0)$ and $\epsilon \equiv l - h_0$. In terms of these parameters,  $h(0) = h_0$  and $l$ are given by asymptotic expansions $h_0 = h_{out}(0) + m_1 \epsilon + o(\epsilon)$,  $l = h_{out}(0) + n_1 \epsilon + o(\epsilon)$ where $n_1 = m_1 +1$  depends on $h_{out}(0)$ but not on $\epsilon$. 

\begin{figure}[!ht]
\begin{center}
\includegraphics[width=0.9 \linewidth]{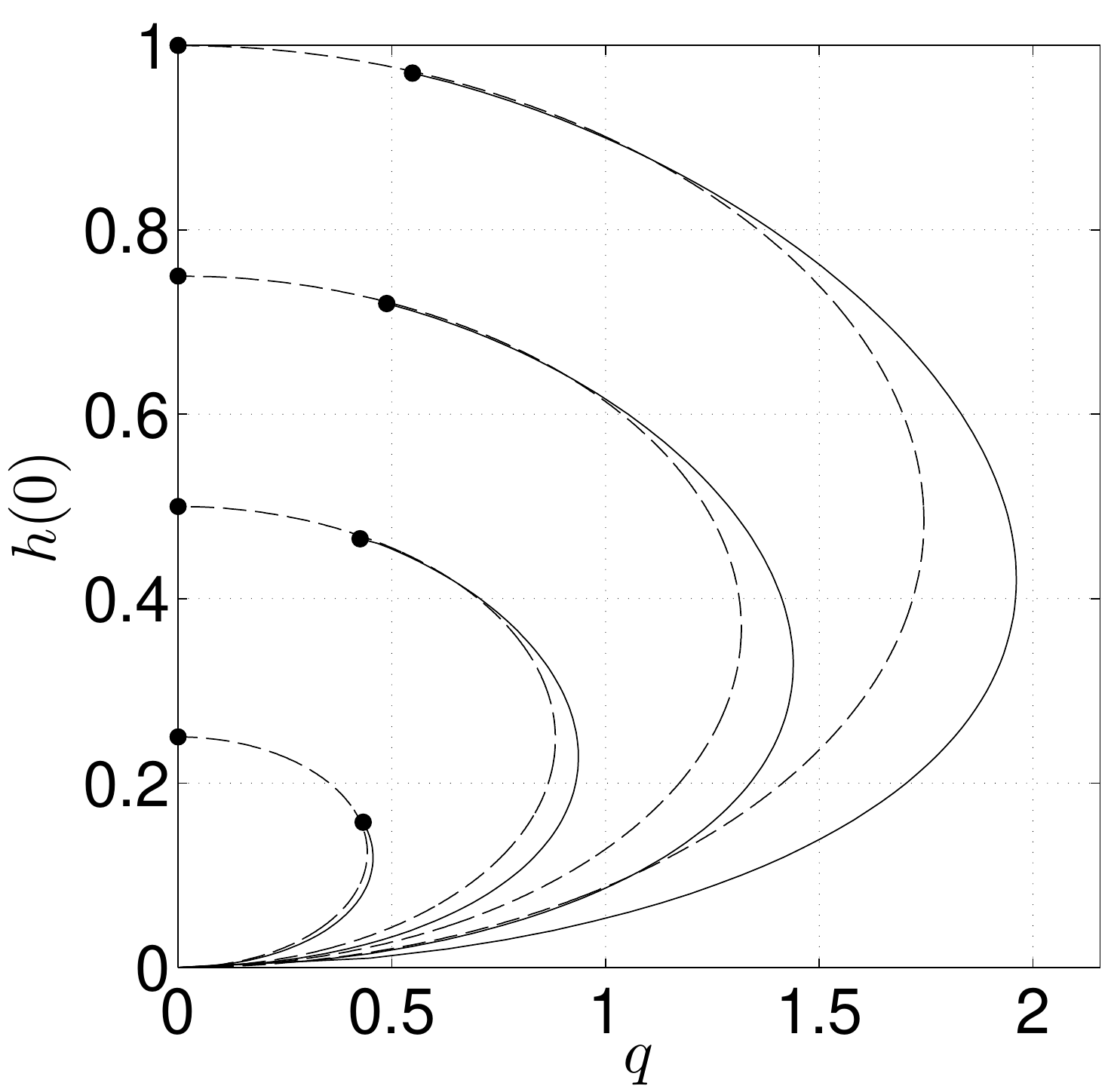}
\caption[Example bifurcation diagrams produced using the collocation method of \S\ref{--S:elec_forward-colloc} compared to those produced using the leading order large deflection approximation of \S\ref{--S:elec_match-pt-outer}]{\label{f:pt_leading_h_vs_q_small_l} The bifurcation diagrams of Fig.~\ref{fig:orig_bifn_diags} (solid curves) compared to diagrams based on leading order large deflection profiles (dashed curves) with $l_0 \in \lbrace 0.25, 0.5, 0.75, 1 \rbrace$. The leading order   solutions give  consistent extensions of the computed bifurcation diagrams to small-$\epsilon$ solutions, as well as capture the overall saddle-node bifurcation structure of the system. Unsurprisingly, the accuracy is generally poor when $\epsilon/l$ is $O(1)$.}
\end{center}
\end{figure}

Combining Eqs.~\eqref{force},~\eqref{e:pt_out_ode_1st_int}~and~\eqref{eq:jump}, we obtain
$$
q \approx \sqrt{\left(2\pi h_{out}(0) \sqrt{4-(h_{out}(0))^2}\right)(l-h_0)}
$$
for fixed $l$. In order to compare this expression with the numerical result in Fig.~\ref{fig:orig_bifn_diags}, we need to compute (or estimate) $h_{out}(0)$ for a given $h_0$ and $l$. As we argue above 
$h_{out}(0) - h_0$ 
is $O(\epsilon)$, so we get
\begin{align}
q & = \sqrt{\left(2\pi h_0 \sqrt{4-h_0^2}\right)(l-h_0)} + O(\epsilon^{3/2}) \label{leading_order_bifn_relationship} 
\end{align}
This relationship fully describes the leading order bifurcation diagram behavior $q \sim \sqrt{l-h_0}$ in the small-$\epsilon$ regime where numerical methods face the greatest challenges. Figure \ref{f:pt_leading_h_vs_q_small_l} compares a typical computed bifurcation diagram from Fig.~\ref{fig:orig_bifn_diags} to the corresponding leading order large deflection diagram based on \eqref{leading_order_bifn_relationship}. The leading order approximation is seen to provide a consistent extension of the computed bifurcation diagram to the small $\epsilon$ regime and captures the overall bifurcation structure of the system even for $O(1)$ values of $\epsilon$.

\subsection{Outer conformal maps} \label{sec:outer}

We now outline a new matching scheme based on numerical conformal mappings that combines high accuracy for all values of $\epsilon$ with efficient handling of the small $\epsilon$ regime. The scheme modifies leading order large deflection profiles with a more subtle description of the true tip shape. We adopt the usual notation of matched asymptotics from this point forward, labeling the conformal maps that describe leading order large deflection profiles `outer' solutions, and the conformal maps that represent the sheet tip profile `inner' solutions.

We remark that rather than just the solution profile $h_{out}(x)$ (see \eqref{eq:outer-profile}) we need an appropriate conformal parameterization $ig(\psi) = x(\psi) + i h_{out}(x(\psi))$ for the outer region.  The use of conformal maps to represent both inner and outer solutions is essential for maintaining the fundamental relationship between global sheet geometry and electric field that characterizes this problem, {\em i.e.} the localization of the force balance in \eqref{eq:new_ode}.

In the same complex-variable framework as in the original numerical scheme, outer solutions are governed by the equation
\begin{equation}
- \text{Re}[g(\psi)] + \frac{\text{Im}[g_{\psi\psi}(\psi)\overline{g_{\psi}}(\psi)]}{\vert g_{\psi}(\psi) \vert^3} = 0, \label{outer_ode}
\end{equation}
where $\lbrace i g(\psi): \psi \in (-\pi,\pi]\rbrace$ is the corresponding deflection profile. This is obtained from \eqref{eq:new_ode} by dropping the electrostatic pressure term, and it is indeed Eq.~\eqref{elastic} away from $x=0$. 

The relevant quantities in \eqref{outer_ode}, namely the deflection of the interface and its curvature, are properties of the profile $h_{out}(x)$ but are independent of the particular parameterization $g(\psi)$ of the interface, with the restriction that $\psi = 0$ is the tip and $\psi \to \pm \pi$ correspond to the boundary condition $h_{out}(x) \to 0$ as $x \to \pm \infty$. Thus, if a conformal map $G$ on $D$ has boundary values $g(\psi)$ which satisfies \eqref{outer_ode}, then so will the composite map $g\circ \mathcal{A}$ where $\mathcal{A}(w)$ is any conformal automorphism from the closed unit disc $D \cup \partial D$ to itself \cite[Chapter 8]{SteinShakarchi} that fixes the points $w=1$ and $w=-1$. Since all conformal automorphisms defined on $D \cup \partial D$ also map the unit circle $\partial D$ to itself, these compositions alter the parametrization but not the shape of the original deflection profile.  This symmetry of the equation is akin to ``gauge freedom" in other physical systems, where one is allowed to make certain transformations which keep all the ``physical" quantities (here the deflection profile) invariant. The collection of conformal automorphisms of $D \cup \partial D$ fixing $1$ and $-1$ forms a one parameter group of mappings 
$$
\mathcal{A}_a(w) = \frac{w-a}{1-aw} \text{ for }a \in (-1,1).
$$ 
with the group composition law 
$$
\mathcal{A}_a \circ \mathcal{A}_b = \mathcal{A}_c \text{ where } \frac{1-a}{1+a}  \cdot \frac{1-b}{1+b} = \frac{1-c}{1+c}.
$$

We use a collocation method, as in sec~\ref{sec:collocation} to compute a numerical conformal map that (approximately) solve \eqref{outer_ode}. An important consideration in developing a collocation method to solve for a conformal map, is to first construct a finite dimensional family of conformal maps, that is {\em adapted} to the particular problem we are trying to solve. In particular, the family of maps in \eqref{eq:discretize} are appropriate for interfaces which decay to zero and are otherwise smooth. This family is therefore {\em not appropriate} for seeking solutions of the outer equation~\eqref{outer_ode}. While the outer profile $h_{out}(x)$ still decays as $x \to \pm \infty$, the profile is no longer smooth -- It has a corner with an outer angle $\pi/\eta$ at $x = 0$. 

In order to incorporate this corner with a know angle, we consider a family of conformal maps given by the composition of the family of smooth maps in~\eqref{eq:discretize} with a fixed analytic map which generates a corner. 

We define the ``corner map" $C_{\eta}$ by
\begin{equation}
C_{\eta}(\zeta) = \frac{\zeta^{\frac{1}{\eta}}}{(\zeta+t)^{\frac{1}{\eta} -1}} + h_{out}(0)
\label{corner-map}
\end{equation}
where $t > 0$ is an arbitrary parameter. This map is analytic and single valued except on a branch cut which we can choose to be segment $[-t,0]$. Also, $C_\eta(0) = h_{out}(0)$, and the image of the imaginary axis under the map $C_\eta$ will have a corner at $h_{out}(0)$ with the correct ``outer" tip angle $\pi/\eta(h_{out}(0))$. Finally, we emphasize that the parameter $t > 0$ is arbitrary and a useful numerical check of our method is to verify that our results are independent of the specific choice we make for $t$. 

As we discuss in Appendix~\ref{apndx:outer}, we get an appropriate family of candidate maps for approximating  the outer solution by composing this corner map with the family of maps in \eqref{eq:discretize}.  In particular,   we seek (approximate) solutions of the outer equation within the family of conformal maps
\begin{equation}
\widetilde{G}(w) = C_\eta\left(\alpha \frac{1-w}{1+w} + \sum_{j=0}^{M_{out}} \beta_j w^j - h_{out}(0) -\frac{t(\eta-1)}{\eta}\right)
\label{outer-conformal}
\end{equation}
We numerically determine the constants $\beta_j$ and $\alpha$ that give the outer solution by requiring that the function $\widetilde{g}(\psi) = \widetilde{G}(e^{i \psi})$ satisfies the appropriate normalization and boundary conditions, and \eqref{outer_ode} at the {\em collocation points} $\psi_m = \frac{m \pi}{M_{out}}, m=0,1,2,\ldots,M_{out}-1$. The details of our numerical implementation are presented in the Appendix~\ref{apndx:outer}.

\subsection{Inner conformal maps} \label{sec:inner}

The sharp corners of outer solutions are regularized by elastic tension close to the tip as shown in Fig.~\ref{fig:schematic_solns}. Electrostatic pressure and sheet curvature are both high near such geometry, so we anticipate the existence of a scaling in which the stabilizing effect of gravity is relatively small. As evident from Fig.~\ref{fig:schematic_solns} and from the fact that the outer solutions give a corner at $x = 0$, the break down of the outer solution occurs on a length scale $\epsilon$ that is set by the separation between the charge location and the interface tip. Thus we set $f = h_{out}(0) + \epsilon \gamma$ (equivalent to scaling all physical lengths by $\epsilon$ but recognizing that the tip is within $O(\epsilon)$ of the prescribed $h_{out}(0)$) and $q = \sqrt{\epsilon} Q$ (motivated by the scaling in \eqref{leading_order_bifn_relationship}) in \eqref{eq:new_ode}. Dropping terms that are $O(\epsilon)$ results in a balance between the electrostatic pressure and surface tension:
\begin{equation}
\frac{Q^2}{4 \pi^2 \vert \gamma_{\theta}(\theta) \vert^2} + \frac{\text{Im}[\gamma_{\theta\theta}(\theta)\overline{\gamma_{\theta}}(\theta)]}{\vert \gamma_{\theta}(\theta) \vert^3} = 0 \label{inner_ode}
\end{equation}
where $\gamma$ and $Q$ are both $O(1)$, and $\lbrace i \gamma(\theta): \theta \in (-\pi,\pi)\rbrace$ is the corresponding rescaled deflection profile near the sheet tip. This equation, along with the condition $\gamma(-\theta) = \overline{\gamma(\theta)}$ has two one-parameter symmetries. We can translate the solutions by a real constant $\gamma \to \gamma+c$, because the equation only sees derivatives of $\gamma$, and we have not as yet imposed any boundary conditions on $\gamma$ which break this translation invariance. The second symmetry is the scaling $\gamma \to k \gamma, Q \to \sqrt{k} Q$ for all $k > 0$. This is related to physical observation that without gravity (and hence without the capillary length scale), the equation should have a scale invariance, as it has no intrinsic length scale. As in the previous section, we therefore do not expect unique solutions to \eqref{inner_ode} without specifying additional {\em normalization(s)} of the solution. And with an appropriate normalization, we expect to find a one parameter family of solutions that are determined entirely by the specified value of $0 < h_{out}(0) <\sqrt{2}$ or equivalently $\eta(h_{out}(0))$, as this was the only input in determining the normalized outer solution.

\begin{figure}[h!]
\includegraphics[width =  0.9 \linewidth]{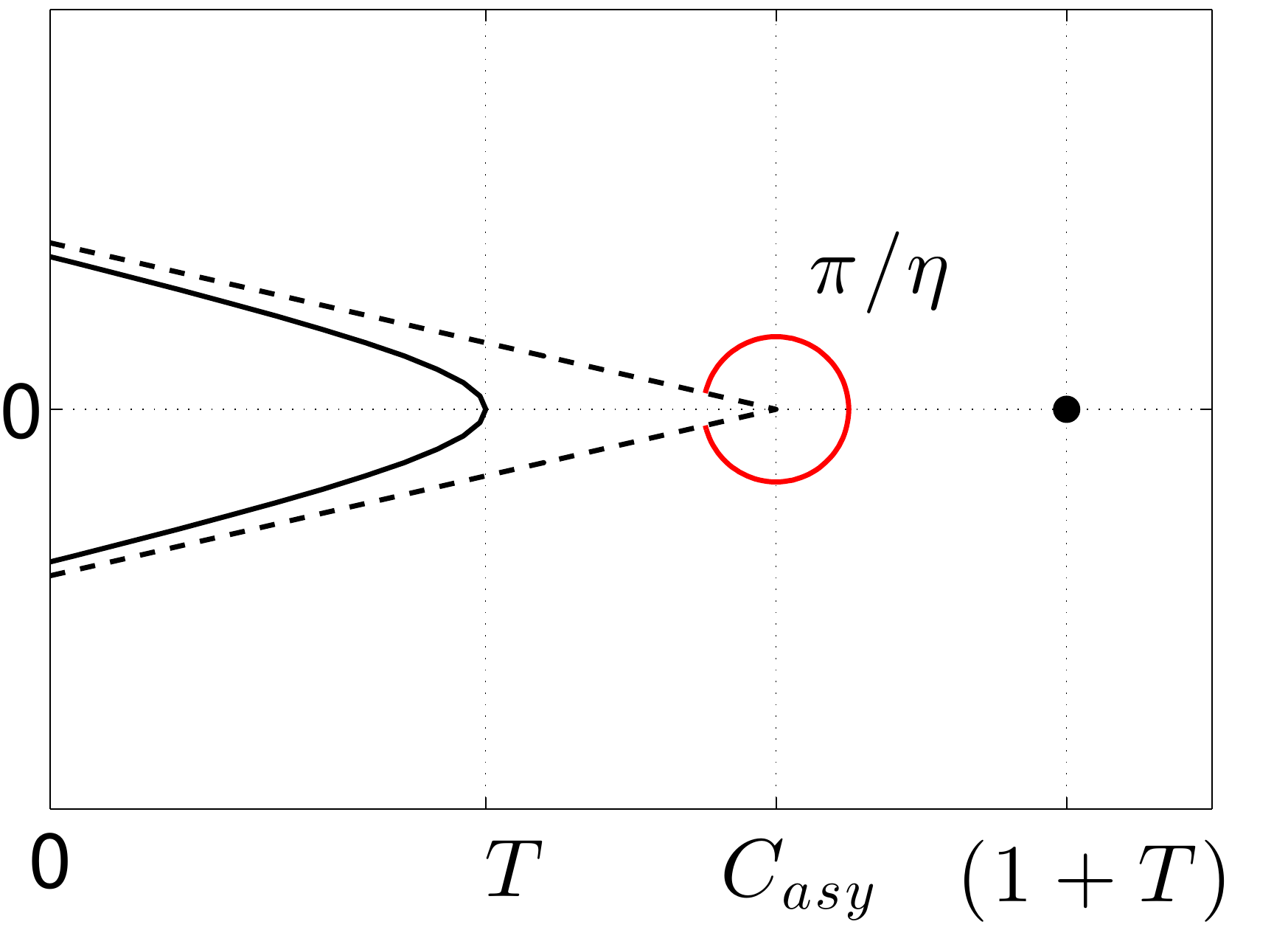}
\caption[Schematic inner solution]{\label{fig:inner_schematic} Schematic rescaled  inner solution map $\Gamma(\omega)$. The maps depend on an arbitrary ``normalization" parameter $T$. A check for our numerical method is that as $T$ is varied, the curves should shift by a corresponding amount in the $x$-direction. This graph is rotated relative to the profiles $(x,h(x))$ since the profile is represented by $ih_0 + i \epsilon \Gamma$. }
\end{figure}

We now discuss the appropriate normalization conditions as well as the discretization of \eqref{inner_ode} in order to numerically compute the inner solutions which describe the tip region. An immediate observation from the form of \eqref{inner_ode}, or \eqref{eq:new_ode} with gravity neglected, is that the curvature is non-positive everywhere. In order to match the outer solutions (which have a corner at 0), it is necessary that as $\theta \to \pm \pi$, the inner profile $\gamma(\theta)$ should approach a pair of straight line asymptotes, that make an (outer) angle $\pi/\eta$. This situation is depicted schematically in Fig.~\ref{fig:inner_schematic}.

Let $\Gamma$ denote the conformal map on the (open) unit disk $D$ whose boundary values on the unit circle determine $\gamma$. We normalize away the scale invariance for $\Gamma$ by setting the separation between the tip of the interface and the charge location at $1$, and incorporate the value of $\eta$ by setting ``boundary conditions" on the behavior of $\Gamma(\omega)$ as $\omega \to -1$: 
\begin{subequations} \label{e:pt_in_bcs_recast}
\begin{align} 
 \Gamma(0) - \Gamma(1) & = 1, \label{e:pt_in_bcs_recast_1} \\
 \Gamma(\omega) -\left[C_{\text{asy}} + \left(A\frac{1-\omega}{1+\omega}\right)^{\frac{1}{\eta}}\right]& \to 0 \text{ as } \omega \to -1 \label{e:pt_in_bcs_recast_2}
\end{align}
\end{subequations}
where $C_{\text{asy}}$ and $A$ are constants whose values are determined as part of the numerical solution. We refer to $C_{\text{asy}}$ as the `asymptotic offset' of a computed solution. We still have the translation symmetry $\Gamma \to \Gamma+ c, C_{\text{asy}} \rightarrow C_{\text{asy}}+c, A \to A, Q \to Q$  for any $c \in \mathbb{R}$. 

For our numerical solutions, we need a family of conformal maps that is adapted to the boundary conditions for the inner solution. As we show in Appendix~\ref{apndx:inner}, an appropriate family of conformal maps for the inner solutions is 
\begin{equation}
\widetilde{\Gamma}(\omega) = \left[A \frac{1-\omega}{1+\omega} + \sum_{j=0}^{M_{in}-1} C_j \omega^j + C_{M_{in}} \left(\frac{1+\omega}{2}\right)^{\frac{1}{\eta}-1}\right]^{\frac{1}{\eta}}
\label{inner-conformal}
\end{equation}
that depends on $M_{in}+2$ real parameters $C_j, j=0,1,2, \ldots,M_{in}$ and $A$. To get a unique solution, we break the translation invariance of the solutions by specifying the additional requirement
\begin{equation}
\Gamma(1) = T \text{ for some } T > 0. \tag{\ref{e:pt_in_bcs_recast}c} \label{e:pt_in_bcs_recast_3}
\end{equation}
The constant $T$ is (somewhat) arbitrary and the restriction $T > 0$ is explained in the appendix. The values $\beta_j$ and $\alpha$ that give the outer solution by requiring that the function $\widetilde{\gamma}(\theta) = \widetilde{\Gamma}(e^{i \theta})$ satisfies the appropriate normalisation and boundary conditions, and \eqref{inner_ode} at the {\em collocation points} $\theta_m = \frac{m \pi}{M_{in}}, m=0,1,2,\ldots,M_{in}-1$. (Details in Appendix~\ref{apndx:inner}).

\begin{figure*}[!ht]
\begin{center}
\subfigure[]
{
\includegraphics[width = 0.45 \linewidth]{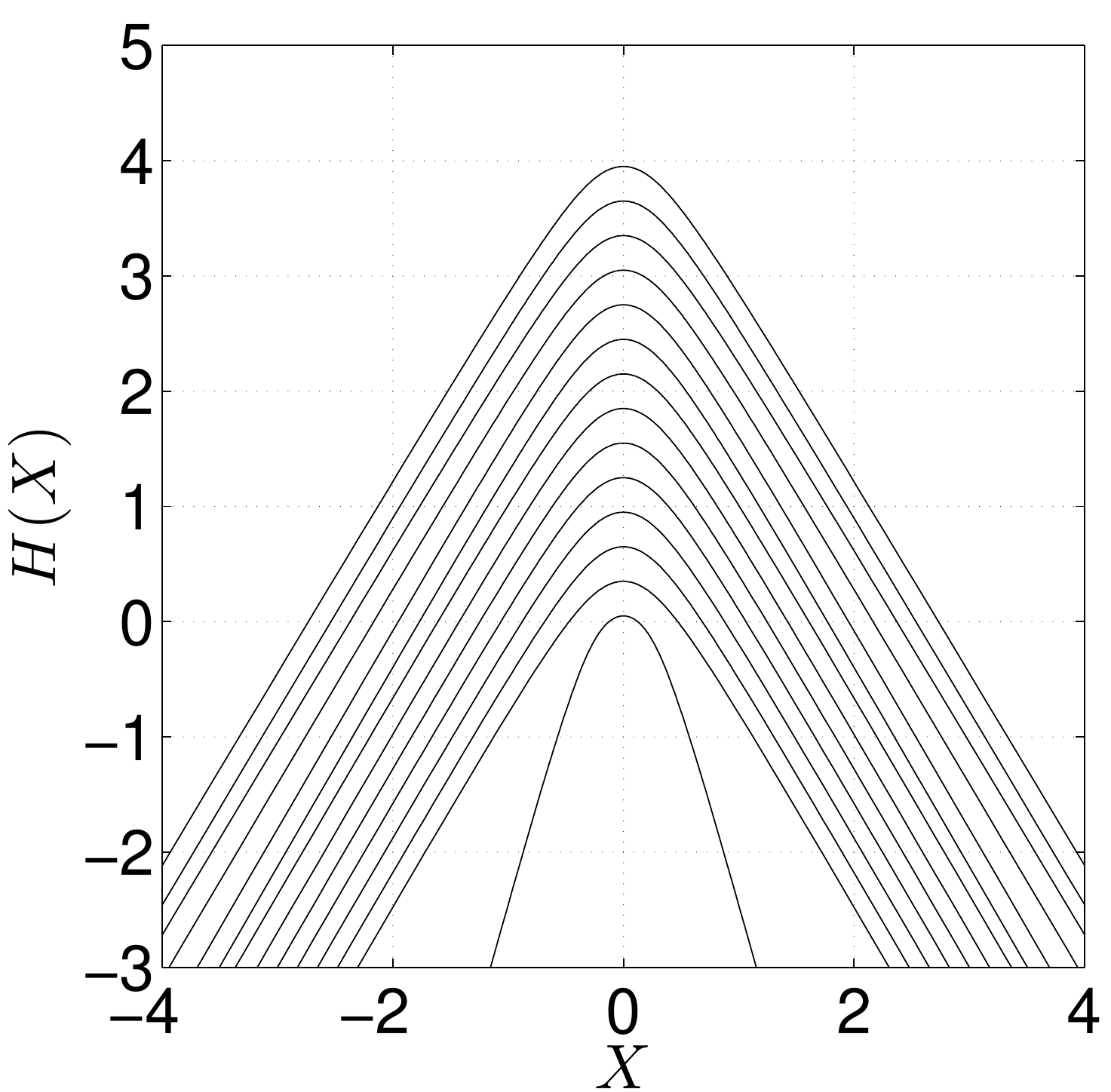}
\label{f:pt_in_comp_prof}
}
\subfigure
{
\includegraphics[width = 0.45 \linewidth]{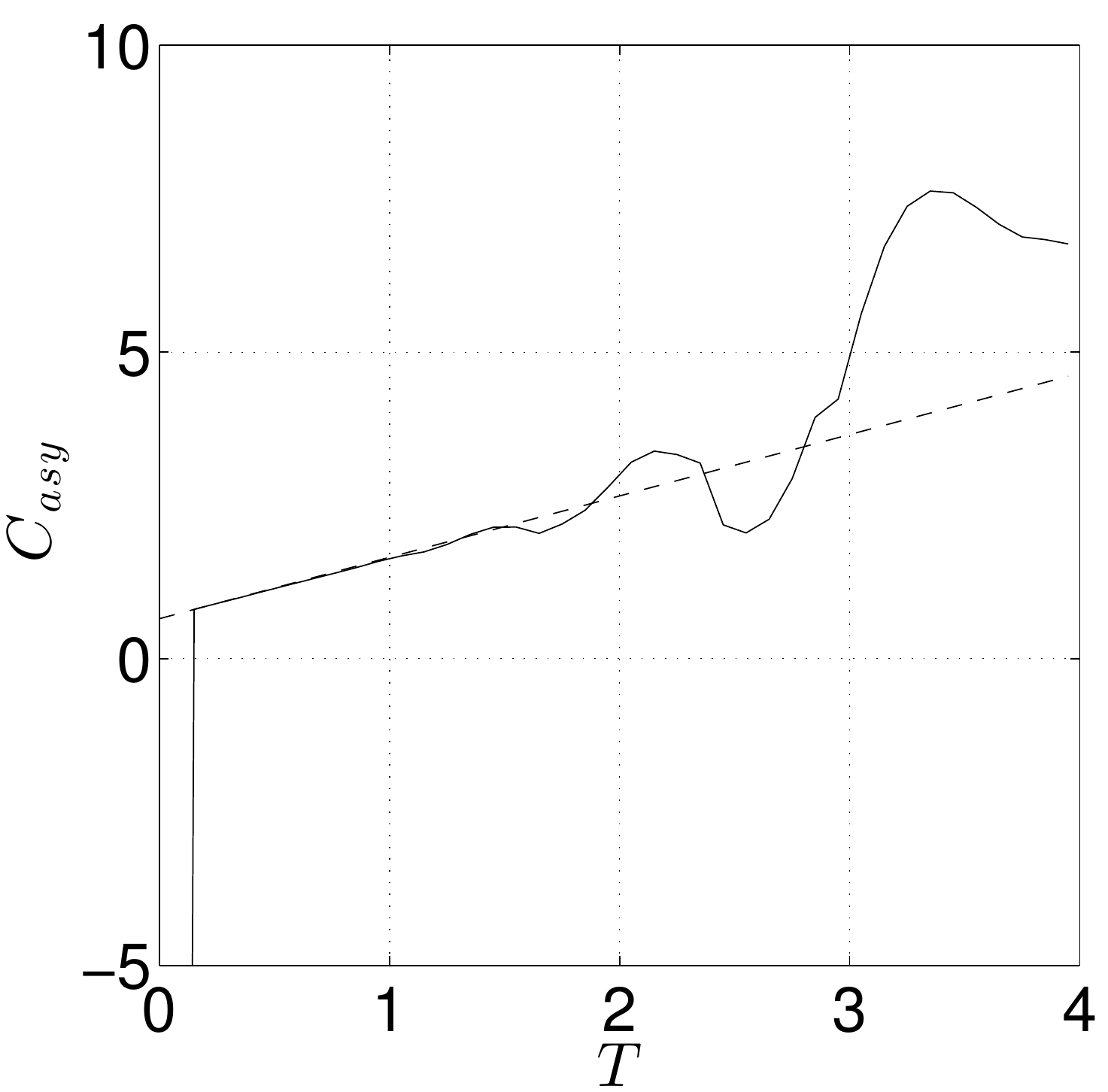}
\label{f:pt_in_offset_vs_T}
}
\caption[Computed inner profiles]{\label{fig:inner_profile} \subref{f:pt_in_comp_prof} Computed inner solution profiles for $\eta = 0.6$, $0.05 \leq T \leq 4$. The profile for $T = 0.05$ is clearly inaccurate; the profiles for $T \gtrapprox 0$ less obviously so. \subref{f:pt_in_offset_vs_T} Asymptotic offsets for the profiles in figure~\ref{f:pt_in_comp_prof} (solid curve), plotted against $T$. The dashed curve shows the theoretical relationship \eqref{e:pt_in_Casy_T_relationship}.}
\end{center}
\end{figure*}

For intermediate values of $\eta$ (not too close to $1/2$ or equivalently $h_{out}(0)$ not close to $\sqrt{2}$) and $T$ moderately close to $0$, the algorithm for computing inner solutions produces curves with all of the desired properties (Fig.~\ref{f:pt_in_comp_prof}). 

A check of our numerical method is that the physical results should be independent of the specific choice of $T$ (as with the choice of the parameter $t$ in computing the outer solutions). For instance, the asymptotic offset $C_{\text{asy}}$ can be computed in terms of the coefficients in the map $\widetilde{\Gamma}$ as $C_{\text{asy}} = \frac{1}{\eta}C_{M_{in}} A^{\frac{1}{\eta}-1}$. The symmetries of the equation now imply that  we should have the identity
\begin{equation}
C_{\text{asy}}(T) = \frac{1}{\eta}C_{M_{in}} A^{\frac{1}{\eta}-1} = T + k \label{e:pt_in_Casy_T_relationship}
\end{equation}
for some constant $k$ depending only on $\eta$ and independent of $T$. This relation is not respected by computed solutions if $T$ becomes either very close to $0$ or larger than a  (\emph{a priori} unknown) threshold value (Fig.~\ref{f:pt_in_offset_vs_T}). For the value of $\eta$ used in these demonstrations, the expected linear behavior is only displayed for $0.1< T < 1$. In addition to giving a check on our numerical method, computing profiles for several $T$ values allows us to identify the range that produces consistent solutions.

\subsection{Matching: Uniformly valid composite conformal maps} \label{sec:matching}

Now that we can compute inner and outer solutions separately, we need to combine them to produce an approximate solution of the full system \eqref{eq:new_ode}.  This is not straightforward, as conformal maps are ``rigid" global objects. Conformal maps on the disk are necessarily smooth (infinitely differentiable), and the real and the imaginary parts must satisfy the nonlocal relation \eqref{eq:constraint}.  Consequently, we cannot simply patch together two conformal maps on different portions of the disk, to produce a composite conformal map.  In fact, if two conformal maps agree in a small neighborhood of a single point, then by analytic continuation they necessarily are identical, i.e. they have a common maximal domain of analyticity and the maps are identical on this domain \cite{Ahlfors-book}. Since the inner and the outer conformal maps are not identically equal (for instance near the tip), they are different ``everywhere", and any map that agrees with one cannot agree with the other. At first sight one might conclude there {\em there is no way to put two distinct conformal maps together}. 

Of course, we do not require that the composite map agree {\em exactly} with either the inner or the outer conformal map. Rather the combined map should approximate the inner and the outer conformal maps ``well" on different pieces of the unit disk. We will now turn this qualitative notion into a quantitative algorithm for combining inner and outer conformal maps to generate a composite map to solve \eqref{eq:new_ode} with {\em a uniformly small error}.

 As discussed above the outer equation \eqref{outer_ode} and the inner equation \eqref{inner_ode} have nontrivial symmetries. In order to compute (unique) numerical solutions of these equations, we break these symmetries by imposing additional normalization conditions. We now have to ``un-normalize" i.e., pick the appropriate symmetry transformed versions of the computed inner and the outer solutions, so that, together they give a solution for the original system \eqref{eq:new_ode}.  
This is achieved by the {\em matching principle}, which requires that there exist an overlap region where both the inner and outer solution describe the true physical solution, and thus agree with each other {\em asymptotically} \cite{vanDyke1975,Hinch1991}.

The full (symmetry related) family of conformal maps describing the inner solution is given by $\widetilde{F}(\omega) = h_{out}(0) + \epsilon(\widetilde{\Gamma}(\omega) + c)$, where $\omega \in D$ is the {\em inner variable} and  $\epsilon > 0,c \in \mathbb{R}$ are elements of the symmetry group, describing scaling and vertical translation of the inner solution. The ``outer limit" of this inner solution is the behavior of this map as $\omega \to -1$ and from our representation for the inner solution a direct calculation yields (see also \eqref{e:pt_in_bcs_recast_2}) 
\begin{align*}
\widetilde{F}_{in}(\omega)  = & h_{out}(0) + \epsilon \left[\left(A\frac{1-\omega}{1+\omega}\right)^{\frac{1}{\eta}} +\frac{C_{M_{in}} A^{\frac{1}{\eta}-1}}{\eta}+c\right]  \\
&  +  \epsilon O((1+\omega)^{\frac{1}{\eta}}).
\end{align*}
The full (symmetry related) family of conformal maps describing the outer solution is given by $\widetilde{F}_{out}(w) = \widetilde{G}(w)$, where the reparameterization invariance of the outer solution implies that the {\em outer variable} $w$ has no intrinsic physical meaning unless it is $1$ (corresponding to the tip) or $-1$ corresponding to the point at $\infty$. Rather if $w$ and $w'$ are two variables related by a conformal automorphism that fixes $\pm 1$,
$$
w' = \mathcal{A}_a(w) = \frac{w-a}{1-aw}, \quad a \in (-1,1)
$$
then both $\widetilde{G}(w') = \widetilde{G
}\circ \mathcal{A}_a(w)$ and $\widetilde{G}(w)$ are valid descriptions of the outer solution. 

In contrast, the inner variable $\omega$ does have intrinsic physical meaning ($\omega = 0$ at the charge location, for instance). Thus, fixing the symmetry element in the reparametrization group is the same as determining $a \in (-1,1)$ such that the inner solution $\widetilde{F}_{in}(\omega)$ matches with the outer solution $\widetilde{F}_{out}(\omega) = \widetilde{G}(\mathcal{A}_a(\omega))$ in an overlap region. We write this relation in shorthand as ``the" outer variable $w = \mathcal{A}_a(\omega)$, meaning the outer variable is identified with that particular choice of $w$ for which the matched outer solution is given by the normalized map $\widetilde{G}$ computed numerically as in section~\ref{sec:outer}.

A direct calculation reveals that for $w = \mathcal{A}_a(\omega)$ 
$$
\frac{1-w}{1+w} = \frac{1+a}{1-a} \cdot \frac{1-\omega}{1+\omega}.
$$
so to match the inner and outer solutions,  it is convenient to represent the outer limit of the inner solution and the inner limit of the outer solution in terms of $(1-\omega)/(1+\omega)$ and $(1-w)/(1+w)$ respectively. 

Van Dyke's matching rule \cite{Hinch1991} requires that the outer limit of $\widetilde{F}_{in}$ agree with the inner limit of $\widetilde{F}_{out}$. We can implement this rule by expanding $\widetilde{F}_{in}$ in $(1+\omega)$, $\widetilde{F}_{out}$ in $(1-w)$ and using the relation $w = \mathcal{A}_a(\omega)$ between $w$ and $\omega$. This yields the conditions
\begin{align}
c & = - \frac{C_{M_{in}} A^{\frac{1}{\eta}-1}}{\eta} \label{eq:matching_1} \\
\epsilon^\eta A & = t^\eta \frac{\left(\alpha - 2 \sum_{j=0}^{M_{out}}j \beta_j\right)}{t} \cdot \frac{1+a}{1-a} \label{eq:matching_2}
\end{align}
The quantities $A,\alpha,\beta_j$ and $C_{M_{in}}$ are defined in \eqref{outer-conformal} and \eqref{inner-conformal}. They are obtained by solving problems \eqref{outer_ode} and \eqref{inner_ode} that are {\em independent} of $\epsilon$. Consequently they are $O(1)$ quantities which can only depend on $h_{out}(0)$ and not on $\epsilon$. Matching thus gives the requirement 
$$
\frac{1+a}{1-a} = \left(\frac{\epsilon}{t}\right)^\eta \frac{At}{\left(\alpha - 2 \sum_{j=0}^{M_{out}}j \beta_j\right)} \equiv K \epsilon^\eta 
$$
where we have defined an $O(1)$ constant $K$. The overlap function is calculated using \eqref{eq:matching_1} in $\widetilde{F}_{in}$ to be 
\begin{align}
 \widetilde{F}_{c}(\omega)  = & h_{out}(0) +  \epsilon \left(A\frac{1-\omega}{1+\omega}\right)^{\frac{1}{\eta}} \nonumber \\ 
 = &  h_{out}(0) +t \left(\frac{\alpha - 2 \sum_{j=0}^{M_{out}}j \beta_j}{t}\right)^{\frac{1}{\eta}}\left(\frac{1-w}{1+w}\right)^{\frac{1}{\eta}}, \label{eq:overlap}
\end{align}
From the success in applying Van Dyke's matching rule it follows that
\begin{enumerate}
\item In the ``outer" region $|1+\omega| \ll 1$, the inner solution agrees very well with the overlap function,
\item In the ``inner" region $|1-w| \ll 1$, the outer solution agrees very well with the overlap function,
\item There is a non-trivial overlap region where $|1-w| \ll 1$ and $|1+\omega| \ll 1$. In this region, all the three functions $\widetilde{F}_{in}, \widetilde{F}_{out}$ and $\widetilde{F}_{c}$ agree with each other,
\end{enumerate}
where by ``agree well" we mean that the leading terms in an asymptotic expansion in $\epsilon \to 0$ are identical. This argument can also be recast in terms of {\em matching through an intermediate variable} \cite{Hinch1991}. For any given $0 < r < \eta$ and complex number $\tau$ with $\text{Re}(\tau) > 0$, there is a sufficiently small $\epsilon_0$ such that for all $\epsilon < \epsilon_0$, $-1+\epsilon^r \tau \in D$, the open unit disk. From the relation \eqref{eq:matching_2} between $w$ and $\omega$, we get
\begin{align}
& \text{Re}(\tau) > 0, \quad 
 \omega = -1+\epsilon^r \tau \nonumber \\
 \implies & \quad w = \mathcal{A}_a(\omega) = 1-\frac{4 K \epsilon^{\eta-r}}{\tau} + O(\epsilon^\eta).
\label{eq:intermediate}
\end{align}
Substituting these expressions in $\widetilde{F}_{in}$ and $\widetilde{F}_{out}$ respectively, we get 
\begin{align*}
\widetilde{F}_{in} & =  h_{out}(0) +  \epsilon^{1-\frac{r}{\eta}} \left(\frac{2A}{\tau}\right)^{\frac{1}{\eta}} + O(\epsilon^{1-\frac{r}{\eta} + 2 r})\\
\widetilde{F}_{out} & =  h_{out}(0) +  \epsilon^{1-\frac{r}{\eta}} \left(\frac{2A}{\tau}\right)^{\frac{1}{\eta}} + O(\epsilon^{1-\frac{r}{\eta} + (\eta- r)})\\
\end{align*}
so the first two terms in the asymptotic expansions of the  inner and outer conformal maps agree as $\epsilon \to 0$ with $\tau$ fixed. 
 
\begin{figure*}[!ht]
\begin{center}
\subfigure[]
{
\includegraphics[height=2.5 in]{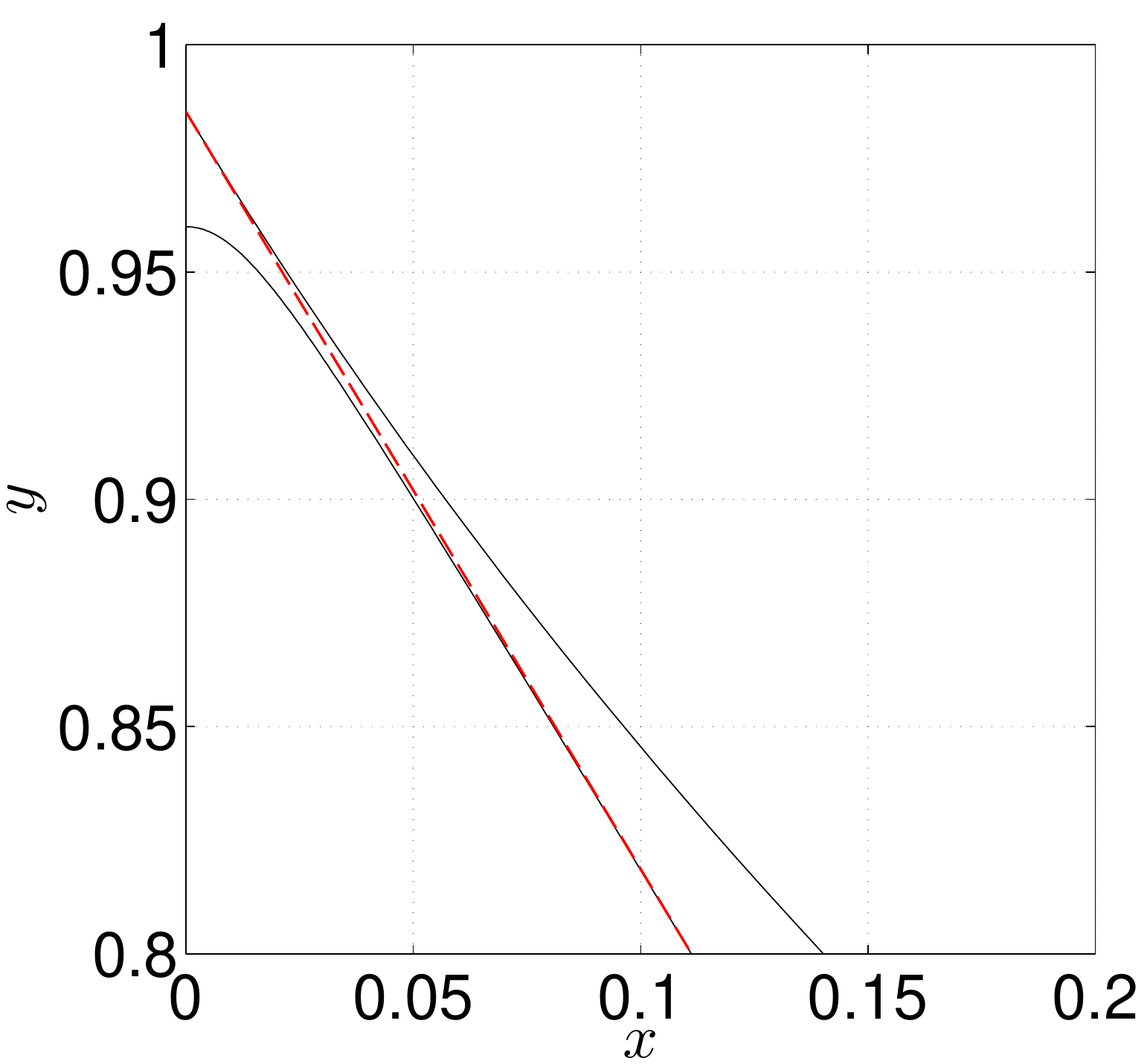}
\label{f:pt_mat_overlap_region1}
}
\subfigure[]
{
\includegraphics[height=2.5 in]{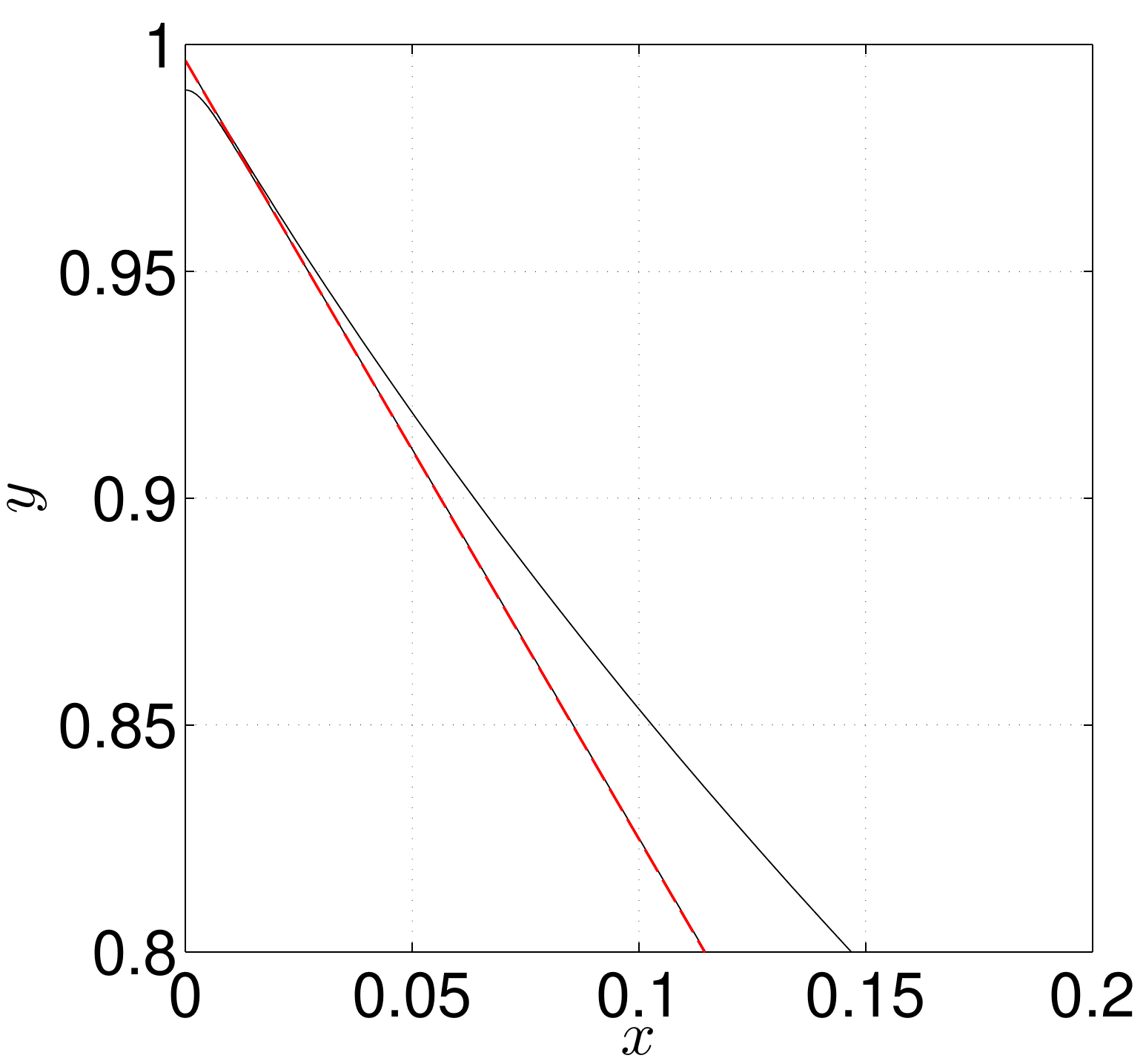}
\label{f:pt_mat_overlap_region2}
}
\label{f:pt_mat_overlap} 
\caption[Plots showing the overlap regions for two inner/outer profile pairs]{\label{f:pt_map_overlap} Inner and outer solution profiles (solid lines) for $l_1 = 1$, $M_{out} = 128$,  $M_{in} = 32$. Each profile is compared to the overlap profile given by the dashed line. By comparing the profiles computed using \subref{f:pt_mat_overlap_region1} $\epsilon = 0.04$ and \subref{f:pt_mat_overlap_region2} $\epsilon = 0.01$, we observe that the region in which both inner and outer profiles agree with the overlap profile to within a given tolerance is of the form $\epsilon k < x < K$ where $k$ and $K$ are both $O(1)$.}
\end{center}
\end{figure*}

  Using $\omega \sim O(1)$ and $\omega \sim -1 + \epsilon^{\eta} \tau$ as the ``boundaries" of the overlap region, we determine that the inner and the outer profiles agree in ``physical" space for $\epsilon \ll x \ll 1$. This makes good physical sense because the the overlap region should be all the length scales between the tip curvature scale where electrostatics balances curvature and the capillary length where gravity balances curvature. In this intermediate regime curvature dominates both gravity and electrostatics, so the solutions correspond to $\kappa = 0$, i.e. straight lines. Figure~\ref{f:pt_map_overlap} shows the overlap region, described by straight lines, where the inner and outer solutions agree.

 We can now construct a composite numerical conformal map $\widetilde{F}$ on $D$ which is a uniformly valid approximation \cite{Hinch1991} (i.e works both in the inner and the outer region) by
 \begin{equation}
 \widetilde{F}(\omega) = \widetilde{F}_{in}(\omega) + \widetilde{F}_{out}(\omega)-\widetilde{F}_{c}(\omega).
 \label{eq:composite}
 \end{equation}
 This map, as a linear combination of conformal maps, is manifestly a conformal map on the unit disk. For $|1+\omega|$ small, i.e. in the outer region $\widetilde{F} \approx \widetilde{F}_{out}$ and for $|1-w|$ small, i.e. in the inner region $\widetilde{F} \approx \widetilde{F}_{in}$. This composite conformal map therefore approaches the right ``limits" asymptotically in the inner and the outer regions (which overlap) and is thus an approximate solution with a uniformly small error. The interface profile is obtained by taking the limit $\widetilde{f}(\theta) = \lim_{\omega \to e^{i \theta}} \widetilde{F}(\omega)$. 

\begin{figure*}[hp]
\includegraphics[width =  0.95 \linewidth]{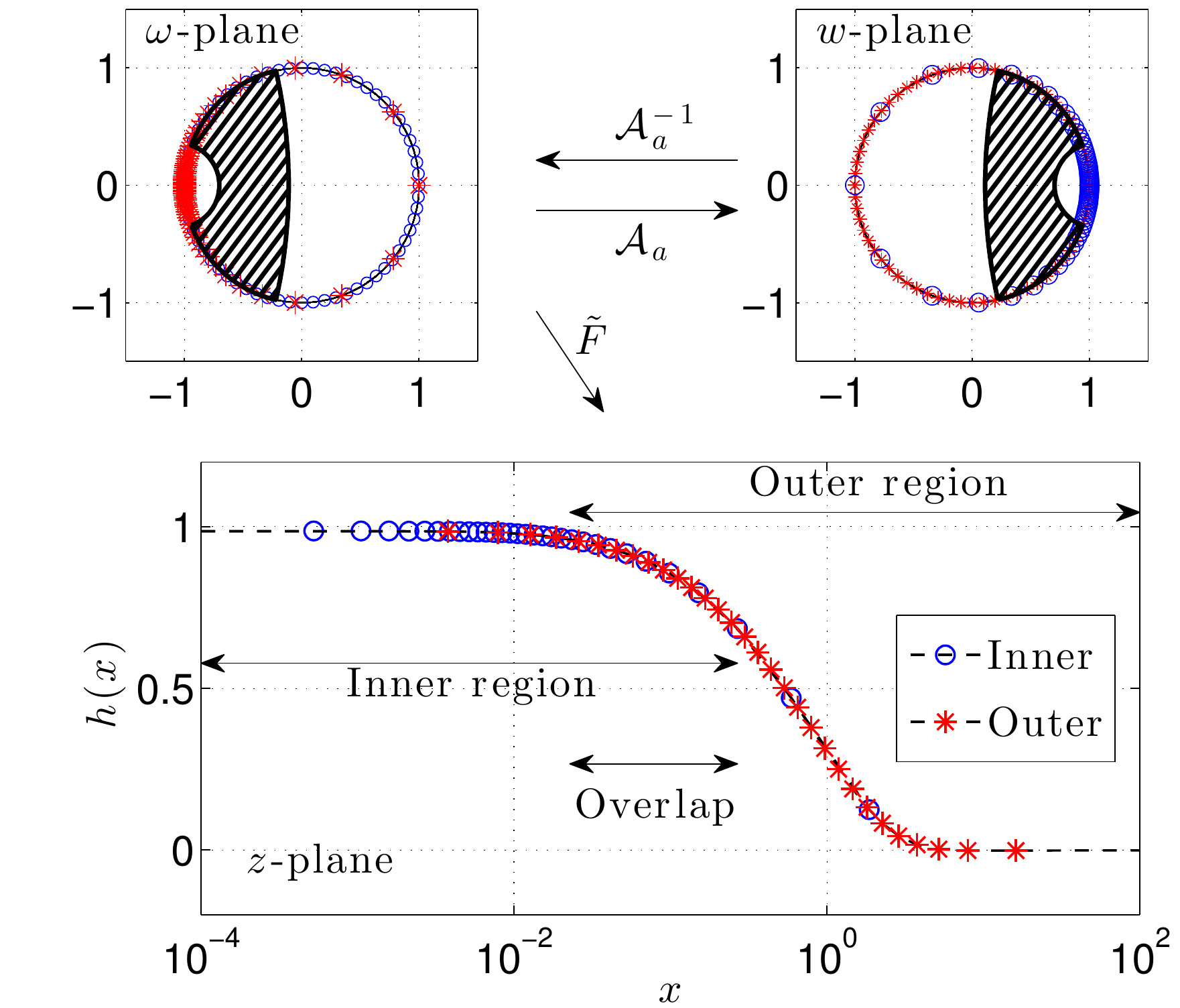}
\caption[Two circles perspective]{\label{fig:two-circles} Schematic representation of the matching procedure. Our matching procedure involves three complex planes, the ({\em inner}) $\omega$-plane, the ({\em outer}) $w$-plane and the ({\em physical}) $z$-plane. The circles $(\bigcirc)$ represent the {\em inner nodes} which are uniformly distributed on the unit circle in the (inner) $\omega$-plane, and the asterisks ({\Large \textasteriskcentered}) are the {\em outer nodes} which are uniformly distributed on the unit circle in the (outer) $w$-plane. They are drawn with different sizes in the $\omega$ and $w$ planes so they can be clearly distinguished. The {\em overlap} region described by the intermediate variable (see \eqref{eq:intermediate}), is hatched in the $\omega$- and the $w$-planes.   The matched profile is plotted on semilog axes to clearly show how the inner nodes (circles) resolve the small $x$ region near the tip while the outer nodes resolve the large $x$ region.}
\end{figure*}

Our matching procedure is schematically illustrated in figure.~\ref{fig:two-circles}. Since we construct the composite map by solving for the inner and the outer conformal maps separately, the underlying domain on which the numerical conformal map $\widetilde{F}$ is defined is naturally interpreted as a {\em pair of (discretized) unit circles}. The outer and the inner variables are related by $w = \mathcal{A}_a(\omega)$ and the matched profile in the (physical) $z$-plane is given by $z = \widetilde{F}(\omega)$ as in \eqref{eq:composite}.  Note the non-uniformity in the relative distribution of the inner an the outer nodes. We emphasize that, for matching conformal maps, the overlap region, as described by the intermediate variable in \eqref{eq:intermediate} is {\em two dimensional}. In particular, our matching procedure requires that the inner and the outer solutions agree on a open set in the complex plane, and it does not suffice, in general, to only have agreement of the inner and the outer profiles, i.e the functions representing $h(x)$. 

On the unit disks in the $\omega$- and $w$-planes, the overlap region is given by the common region between $|1+\omega| \lesssim O(1)$ (the outer region) and $|1-w| \lesssim O(1)$ (the inner region). We indicate this region in the $\omega$- and $w$-planes by hatching in figure~\ref{fig:two-circles}. We remark that the densities of the inner and outer nodes are comparable in the overlap region but not outside it. This is to be expected as the inner and the outer solutions give equally well resolved numerical solutions in the overlap region. For the schematic plot in figure~\ref{fig:two-circles}, the densities do not differ by more than a factor of 2 in the overlap region, so there are no more than 4 and no fewer than 1 circle for every 2 asterisks and vice-versa. In contrast, the nodes outside the overlap region are overwhelmingly either circles (the inner region) or asterisks (the outer region), showing that in these regions either the inner or the outer solution is inaccurate/poorly resolved.

\begin{figure}[ht!]
\includegraphics[width =  \linewidth]{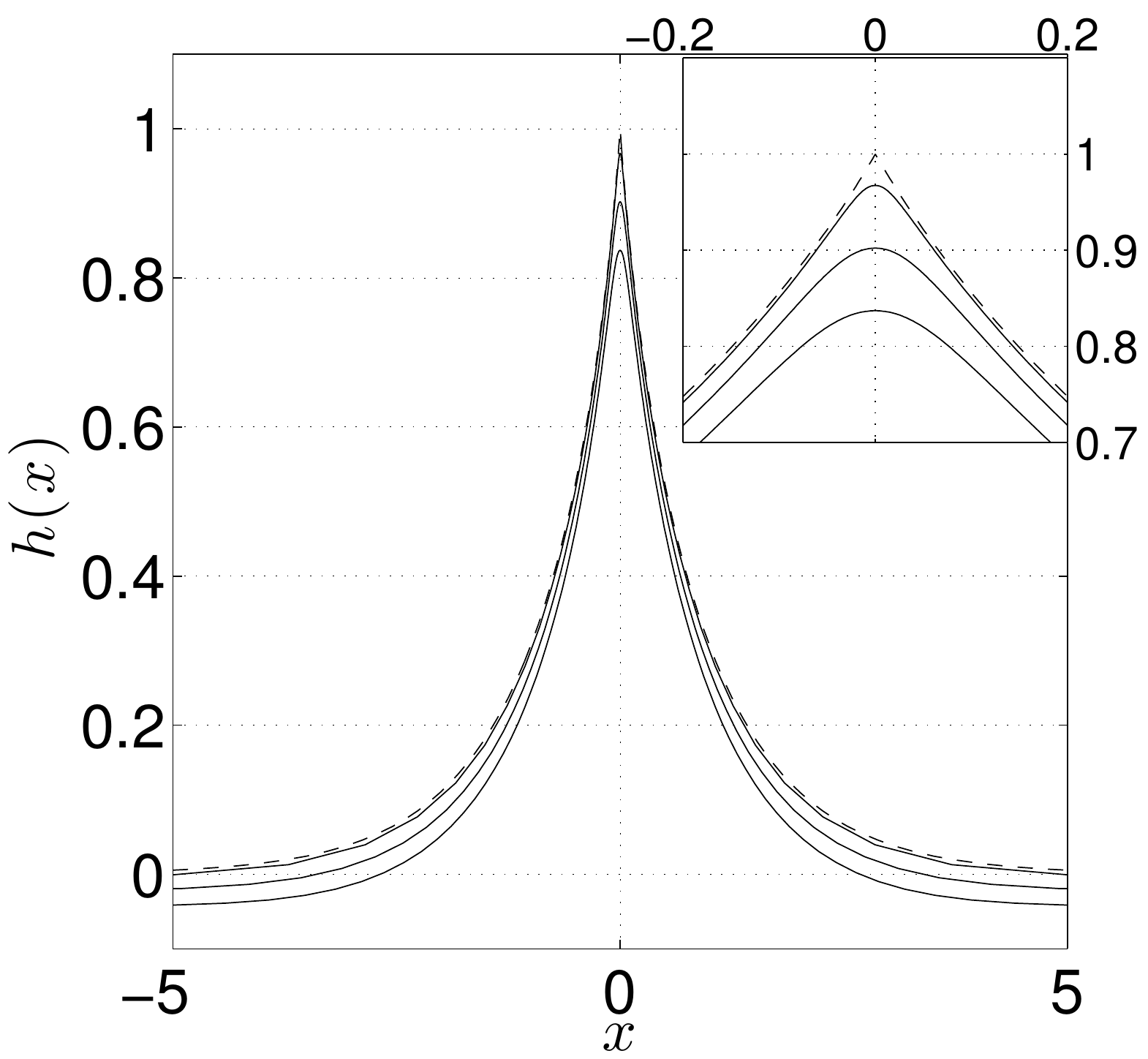}
 \label{f:family_matched_solns}. 
\caption[Families of matched profiles computed by combining a single pair of inner and outer profiles] {Matched profiles computed with $h_{out}(0) = 1$ and $\epsilon = 0.03$, $0.1$, $0.17$. Te dashed curve is the outer solution and the solid curves are the matched solutions. The inner and outer solutions only need to be calculated once to produce these plots. The inset shows the tip shapes for the profiles. The rescaling of the inner solution by $\epsilon$ produces the appropriate tip curvature in each case.}
\end{figure}

Because the inner and outer solutions have their own set of nodes, it suffices to have as few as 16 nodes in each region to get fully resolved matched solutions for the interface. The solution in Figure~\ref{fig:two-circles} was generated with 32 inner and 32 outer nodes.  Figure\ref{f:family_matched_solns} show matched interface profiles created by combining a single pair of inner (32 node) and outer (128 node) profiles using different values of $\epsilon$. Generating profiles in this way (with fixed $h_{out}(0)$) is extremely efficient once the initial calculations required to determine the appropriate inner and outer solutions have been performed. Note that the charge location $l$ is not equal for all these interfaces; rather it varies with $\epsilon$. This is a minor drawback when seeking solutions of the forward problem. In order to compute the solutions with a specifies $l$ and $h_0$, we need to invert the (nonlinear!) relation between these parameters, and the asymptotic parameters $h_{out}(0)$ and $\epsilon$.

Observe that the matched profile $h(x) < 0$ for $x \gtrsim 1$ (a few times the capillary length) in figure~\ref{f:family_matched_solns}. This disagrees with the physical intuition that the force from the line charge will only deflect the interface upwards. In fact, we can prove rigorously that the sheet cannot deflect below the $x$-axis for the system \eqref{e:h_gov_pt}. This dip below the axis in our matched solutions, albeit very small, reveals a possible flaw in our matched solution and therefore demands an explanation. 

In our outer solutions, the balance is between gravity and elasticity, and for large $x$, the profile $h_{out}(x)$ decays exponentially, as in \eqref{eq:outer-outer-profile}. For any multipole source, the  electric field on an asymptotically flat conductor decays algebraically. Far away from the tip, the combination of the forcing charge and the induced charge on the tip is effectively a dipole of strength $q \epsilon \sim \epsilon^{3/2}$. Consequently, the induced charge decays no faster than $\epsilon^{3/2}x^{-3}$ for large $x$, and the electrostatic pressure decays as the square of the induced charge $\sim \epsilon^{3} x^{-6}$. Comparing this expression with the hydrostatic pressure in $h_{out}(x)$ we get a crossover when $\epsilon^{3} x^{-6} \sim e^{-x} \implies x \sim \log(\epsilon^{-1})$. Consequently, no matter how small is $\epsilon > 0$, there always exists an outer-outer region where the dominant balance is no longer between gravity and elasticity. Rather, it is between gravity and electrostatics. Further, the boundary of this outer-outer region, $x \sim \log(\epsilon^{-1})$ is not very different from the capillary length scale $x \sim O(1)$ unless $\epsilon$ is exceedingly small.

A formal calculation keeping higher orders in our composite matched solution confirms the conclusions from this heuristic argument. The outer-outer region is given by the crossover $\exp(-\epsilon^\eta/(1+ \omega)) \sim \epsilon |1+\omega|^{2-\frac{1}{\eta}}$ so that for 
$$
|1+\omega| \lesssim -\frac{\epsilon^\eta}{\log(\epsilon)},
$$ 
it is no longer true that the presumed dominant term $\widetilde{F}_{out}(\omega)$ is larger than the ignored higher order corrections in the presumed smaller term $\widetilde{F}_{in}(\omega) -\widetilde{F}_{c}(\omega)$. In other words, the composite solution in no longer a valid asymptotic expansion in the outer-outer region, and one has to introduce a new ``layer" to resolve the non-uniformity in the expansion of the solution in powers of $\epsilon$ \cite{Hinch1991}. This argument also illustrates one of the strengths of the method of matched asymptotics -- {\em the composite solutions have in them, the information about when they are no longer valid} \cite{Hinch1991}.

As we argued above, the scaling for this outer-outer region, $|1+\omega| \sim \frac{\epsilon^\eta}{\log(\epsilon^{-1})}$ differs from the outer-scale $|1+ \omega| \sim \epsilon^\eta$ only by a logarithmic factor, so that for any realistic $\epsilon$, there is not a clear separation between the outer and the outer-outer regions. This makes the matching problem difficult, as is the case with other matched asymptotics problems that have logarithmic terms \cite{vanDyke1975,Hinch1991}. We will present a more extensive analysis of the outer-outer region and the associated matching problem in a future publication. 

\begin{figure}[th!]
\includegraphics[width = 0.9 \linewidth]{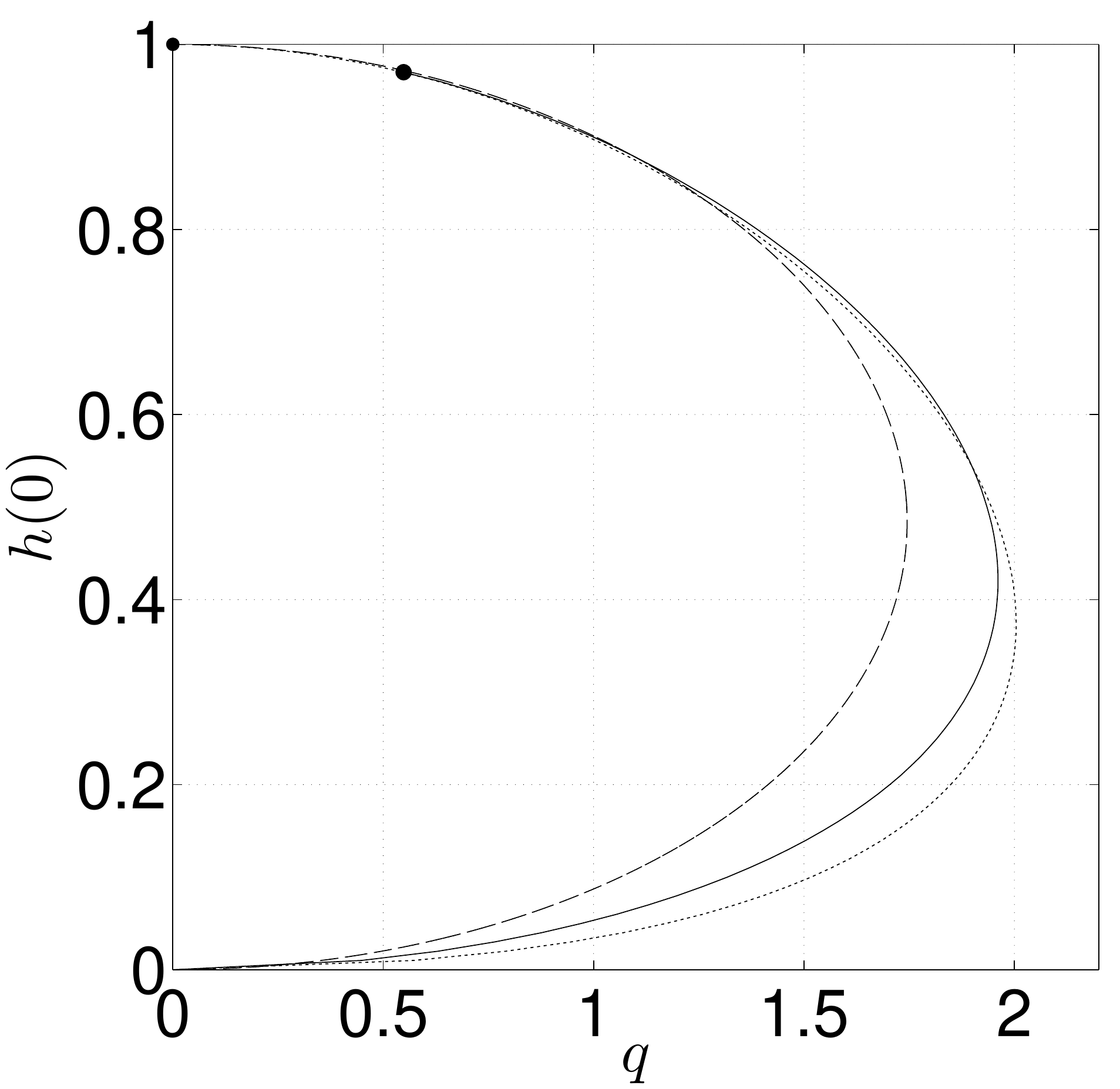}
\caption[Bifurcation diagrams from the original collocation and matching methods]{\label{f:pt_full_vs_lead_vs_mat_bifn}The directly computed (collocation method) and leading order bifurcation diagrams of Fig.~\ref{f:pt_leading_h_vs_q_small_l} (solid and dashed curves respectively) compared to the bifurcation diagram based on matched solutions (dotted curve) with $l_0 = 1$. Note the excellent agreement between the dotted and the solid curves, even for $O(1)$ values of $\epsilon$, despite the matching procedure relying on asymptotic expansions which depend on the smallness of $\epsilon$.}
\end{figure}

We also remark that the outer-outer region has a very small effect on the solution in the inner region near the tip, and on  the computed bifurcation diagram using the matched solution shown in Fig.~\ref{f:pt_full_vs_lead_vs_mat_bifn}.  The figure confirms that the matching technique described in this paper produces uniformly valid, high accuracy solutions (errors are $O(\epsilon)$ in the inner {\em and} the outer regions) for all values of $\epsilon$ while remaining computationally efficient (especially!) when $\epsilon$ is small.

\section{Discussion} \label{sec:discussion}

In this paper, we develop an adaptive numerical conformal mapping method for efficiently computing multiple scale structures in 2 dimensional problems. The central result is that, with our method, we can use conformal mapping techniques to analyze structures that have widely disparate length scales. This is a significant improvement over existing numerical conformal mapping methods for free boundary problems, because they are essentially ``single-scale" and are plagued by the problem of crowding when applied to multiple-scale problems. As we discuss in Sec.~\ref{sec:matching}, there are good reasons why one would not even consider matched asymptotics for conformal maps, and to our knowledge, this is the first work that successfully combines numerical conformal mappings with the method of matched asymptotics.

We consider a model system is governed by three basic forces -- gravity, electrostatics and surface tension. The interface is governed by multiple dominant balances between different forcing/restoring mechanisms, so the free boundary does develop a multiple scale structure. The model problem is thus  a good test bed for developing our method, although it is 
not (to our knowledge) directly relevant to applications.

We reformulate the problem in terms of conformal mappings -- it converts a non-local problem into a problem that is governed by local equations, except some of the equations are in ``real" space an some are in ``reciprocal" space (i.e in the Fourier transform domain). Since the Fast Fourier transform (FFT) allows us to efficiently convert from real to reciprocal space and vice-versa, we can get efficient numerical methods for approximating solutions to our problem in terms of discretized conformal maps. This approach leads to  the collocation method as described in section~\ref{sec:collocation}. 

The FFT requires evenly spaced nodes (in both real and reciprocal space), so we are no longer free to choose the node locations in our discretization. This is a particular concern in multiple-scale problems, since the nodes tend to concentrate in regions of high curvature, leaving the rest of the domain under-resolved. This phenomenon, called  crowding, is a significant impediment to the using conformal mapping methods for multiple-scale problems. The underlying issue, namely the ``stiffness" of the governing equations, also shows up in other methods (boundary integral/boundary element etc.), and is responsible for the challenges of numerically studying free boundary problems with widely disparate scales. In our model, we find that the numerical collocation method no longer converges to a solution once the tip curvature gets sufficiently large.

The ``obvious" physical parameters of the model are the non-dimensional charge strength $q$ and location $l$. However, the ``correct" asymptotic parameters in the sharp-tip regime are the ``uncorrected" tip height $h_{out}(0)$ and the tip-charge separation $\epsilon$. In section~\ref{sec:concentrated} we use physical arguments to determine the charge distribution on the interface and deduce the appropriate scaling for the the physical parameters in terms of the asymptotic parameters. We emphasize that the analysis in this section {\em did not use conformal maps}, and as such the ideas also extend to three dimensional versions the two-fluid or MEMS models. 

Using scalings in terms of the asymptotic parameter $\epsilon$, we reduce the governing equation to an {\em outer} equation that describes the balance between gravity and surface tension, and an {\em inner} equation that describes the balance between electrostatic forces and surface tension. These {\em asymptotic equations} have non-trivial symmetries, and we determine {\em discretized conformal map solutions} of these equations with appropriate normalization using the collocation method as in section~\ref{sec:collocation}. It is important that we pick the discretization appropriately to reflect the known structure of the solutions (a corner in the interface for the outer solution, asymptote to a pair of straight lines with a known angle for the inner solution). Because each region has a single balance, we are not solving for a multiple-scale structure, and the collocation method works well (Sections~\ref{sec:outer}~and~\ref{sec:inner}). 

In applying this method to other problems, it is crucial that we identify the ``correct" asymptotic parameters and scaling relations for the physical quantities, since this knowledge is required to derive the asymptotic limiting equations for the system. We then use the relevant symmetries of the asymptotic equations (The method is powerful precisely because conformal maps have a rich group of symmetries) to determine the appropriate matching conditions that allows us to combine distinct conformal maps on different scales to form a single composite ``multi-scale" conformal map. 

In section~\ref{sec:matching}, we discuss this issue in detail for our model system. The key insight is that the natural domain for discretized multi-scale conformal maps is not a single complex plane. Rather, it is two (or perhaps more for other problems) sets of evenly spaced points on unit-circles, that are related by non-trivial mapping functions (conformal automorphisms of the unit disk). Each discretized unit circle resolves one scale (i.e region with one dominant balance) in the solution, and the requirement that the functions defined on the various circles agree (asymptotically) on their mutual overlaps allow us to patch them together to construct a composite multi-scale conformal map. Since each unit circle is discretized evenly, we can still exploit the efficiency of the FFT in solving the asymptotic equations. and the the morphing of the unit circle by conformal automorphisms allows us to (in effect) pick the nodes in our discretization adaptively and thus resolve structures with disparate length scales in the solution (see figure~\ref{fig:two-circles}).  

This idea seems related to the Cross-ratios of Delaunay triangulation (CRDT) algorithm for  computing Schwarz-Christoffel transformations for nonconvex/multi-armed polygonal domains \cite{CRDT98}. The CRDT algorithm is  designed to combat crowding, and a key idea in is to use multiple maps into unit circles that are related by conformal automorphisms. These automorphisms are chosen  to ``blow-up" portions of the unit circle to ensure that nodes are not crowded (locally). However, unlike our method of asymptotic matching of conformal maps, the CRDT algorithm exploits the invariance of the cross ratio \cite{Ahlfors-book} under conformal automorphisms to (implicitly) construct a composite conformal map. Further work is needed to explicate the precise connection between our multi-scale method for free boundary problems and the CRDT algorithm for given ``multi-scale" polygonal domains.
\section*{Acknowledgements}

This work was supported by the NSF through DMS grant 0807501. We are also grateful to two anonymous referees whose comments on an earlier version helped significantly  improve the presentation and discussion of our results.

\appendix

\section{Outer solutions} \label{apndx:outer}

The outer solutions are entirely determined by a single parameter $h_{out}(0)$. Given a value $0 < h_{out}(0) < \sqrt{2}$, we use \eqref{eq:gamma} to compute the appropriate corner angle $\pi/\eta$. In terms of $\eta = \eta(h_{out}(0))$, we define the ``corner map" $C_{\eta}$ by
\begin{equation*}
C_{\eta}(\zeta) = \frac{\zeta^{\frac{1}{\eta}}}{(\zeta+t)^{\frac{1}{\eta} -1}} + h_{out}(0).
\end{equation*}
with a branch cut on the segment $[-t,0]$. Expanding about $\zeta = 0$ and $\zeta = \infty$, we get 
\begin{align*}
C_{\eta}(iy)  = & h_{out}(0) + \exp\left(\frac{i \, \mathrm{sign}(y) \pi }{2 \eta}\right) \left(\frac{|y|}{t}\right)^{\frac{1}{\eta}} \\
& + O\left(|y|^{1 + \frac{!}{\eta}}\right) \\
C_\eta(\zeta)  = & \zeta+h_{out}(0) +\frac{t(\eta-1)}{\eta}  + O\left(\frac{1}{|\zeta|}\right)
\end{align*}
We define the candidate maps $\widetilde{G}(w) = C_\eta(\widetilde{F}(w)-\widetilde{F}(1))$, where the (generic) maps $\widetilde{F}$ are defined in \eqref{eq:discretize}. The maps $\widetilde{G}$ have the right behaviour at $w = 1$, namely a corner with outer angle $\pi/\eta$ at the point $(h_{out(0)},0)$. As before, the $x \to-x$ symmetry of the outer profile $h_{out}(x)$ implies that the coefficients $\alpha, \beta_0,\beta_1,\ldots $ are all real.

Let $\displaystyle{B(1) = \sum_{j = 0}^{M_{out}} \beta_j}$ and $\displaystyle{B(-1) = \sum_{j = 0}^{M_{out}} (-1)^j \beta_j}$. Using the definition of $\widetilde{G}$ and the behavior of $C_\eta(\zeta)$ as $\zeta \to \infty$, we see that as $w \to -1$,
$$
\mathrm{Re}[ \widetilde{G}]  \to  h_{out}(0)+B(-1) -B(1) + \frac{t(\eta-1)}{\eta}
$$
The boundary condition that $h_{out}(x) \to 0$ as $x \to \infty$ yields
$$
B(1) = h_{out}(0)+B(-1) + \frac{t(\eta-1)}{\eta}.
$$
Finally we have one (real) degree of freedom from the allowed conformal reparameterization of the outer solutions. We fix this with a ``gauge condition" $B(-1) = 0$.

Defining $\psi_m = \frac{m \pi}{M_{out}}, m=0,1,2,\ldots,M_{out}-1$, we have $M_{out}+2$ independent (real) equations, which are respectively the vanishing of the residual at $M_{out}$ equally spaced collocation points, the gauge fixing normalization and the boundary condition as $x \to \pm \infty$. 
\begin{align*}
R[\widetilde{g}(\psi_m)] & \equiv - \text{Re}[g(\psi_m)] + \frac{\text{Im}[g_{\psi\psi}(\psi_m)\overline{g_{\psi}}(\psi_m)]}{\vert g_{\psi}(\psi_m) \vert^3} = 0, \\
B(-1) & \equiv \sum_{j = 0}^{M_{out}} (-1)^j \beta_j = 0 \\
B(1) & \equiv \sum_{j = 0}^{M_{out}} \beta_j = h_{out}(0) +\frac{t(\eta-1)}{\eta}
\end{align*}
This matches up with the number of independent parameters in the candidate conformal parameterization of the interface 
\begin{align*}
\tilde{G}(w) & = C_\eta\left(\alpha \frac{1-w}{1+w} + \sum_{j=0}^{M_{out}} \beta_j w^j - \sum_{j=0}^{M_{out}} \beta_j \right) \\
& = C_\eta\left(\alpha \frac{1-w}{1+w} + \sum_{j=0}^{M_{out}} \beta_j w^j - h_{out}(0) -\frac{t(\eta-1)}{\eta}\right)
\end{align*}

\section{Inner solutions} \label{apndx:inner}

Motivated by \eqref{e:pt_in_bcs_recast_2}, we consider 
$
\Xi(\omega) = \left(\Gamma(\omega)\right)^\eta.
$
A straightforward expansion in $\omega+1$ yields 
$$
\Xi(\omega) = A \frac{1-\omega}{1+\omega} + \frac{\eta C_{\text{asy}}(1+\omega)^{\frac{1}{\eta} - 1}}{(2A)^{\frac{1}{\eta}-1}} + o(|1+\omega|^{\frac{1}{\eta}-1}) 
$$
Recalling that $1/2 < \eta < 1$, we see that even after subtracting the pole at $\omega = -1$, $\Xi(\omega)$ is not analytic at $\omega = -1$ in contrast to the maps in \eqref{eq:representation}. Consequently, in our numerical approximation $\widetilde{\Gamma}$, we need to incorporate this non-analytic behavior as $\omega \to -1$ into the discretization. 

We define a $M_{in}+2$ dimensional family of conformal maps by
\begin{equation*}
\widetilde{\Xi}(\omega) = A \frac{1-\omega}{1+\omega} + \sum_{j=0}^{M_{in}-1} C_j \omega^j + C_{M_{in}} \left(\frac{1+\omega}{2}\right)^{\frac{1}{\eta}-1}.
\end{equation*}
$M_{in}$ is the number of nodes we are choosing to represent the inner solution. $\widetilde{\Xi}$ is a map from the preimage $\omega$ plane to an intermediate plane. The map into the physical plane is then given by $\widetilde{\Gamma}(\omega) = \left[\widetilde{\Xi}(\omega)\right]^{\frac{1}{\eta}}$. The map from the intermediate plane to the physical plane has branch points at $0$ and $\infty$ and we choose the branch cut along the negative real axis. In order that the image $\lbrace \widetilde{\Xi}(e^{i\theta}) : \theta \in (-\pi,\pi) \rbrace$ not intersect this branch cut, we need that $\widetilde{\Xi}(1) =T > 0$.

A conformal parametrization of the interface for the inner solution is given by $\widetilde{\gamma}(\theta) = \widetilde{\Gamma}(e^{i \theta})$. The parameters $A,Q^2$ and $C_j$ for the solution are then determined by the following conditions --
$$
R[\widetilde{\gamma}](\theta_m) =  \frac{Q^2}{4 \pi^2 \vert \widetilde{\gamma}_{\theta}(\theta_m) \vert^2} + \frac{\text{Im}[\widetilde{\gamma}_{\theta\theta}(\theta_m)\overline{\widetilde{\gamma}_{\theta}}(\theta_m)]}{\vert \widetilde{\gamma}_{\theta}(\theta_m) \vert^3} = 0,
$$
is the vanishing of the residual pressure at the collocation points $\theta_m = \frac{m \pi}{M_{in}}, m = 0,1,\ldots,M_{in}-1$;
$$ \sum_{n=0}^{M_{in}-1} (-1)^n C_n = 0
$$
is essentially the boundary condition as $x \to \pm \infty$ since it ensures that 
$$
\widetilde{\Xi}(\omega) - A \frac{1-\omega}{1+\omega} - C_{M_{in}} \left(\frac{1+\omega}{2}\right)^{\frac{1}{\eta} -1} \text{ is } o(|1+\omega|^{\frac{1}{\eta}-1})
$$
so that the boundary condition   \eqref{e:pt_in_bcs_recast_2} holds with $\displaystyle{C_{\text{asy}} = \frac{C_{M_{in}} A^{\frac{1}{\eta}-1}}{\eta}}$; and
\begin{align*} \widetilde{\Xi}(0) & =  A + C_0 + \frac{C_{M_{in}}}{2^{\frac{1}{\eta}-1}} = (1 + T)^\eta, \\\widetilde{\Xi}(1) & =   \sum_{n=0}^{M_{in}-1} C_n + C_{M_{in}} = T^\eta,
\end{align*}
are normalizations for picking a unique element of the two parameter symmetry group of the inner equation \eqref{inner_ode}.

 We thus have $M_{in}+3$ (real) equations to determine $M_{in}+3$ real parameters $Q^2,A$ and $C_j,j=0,1,2,\ldots,M_{in}$ and the constants are determined by minimizing the sum of the squares of the discrepancies in the above $M_{in}+3$ conditions.


%

\end{document}